\documentclass[preprintnumbers,article,amsmath,amssymb,floatfix,11pt,prd,onecolumn,superscriptaddress,nofootinbib,eqsecnum]{revtex4-2}
\usepackage{titlesec}
\usepackage[toc]{appendix}
\usepackage{caption}
\titleformat*{\section}{\LARGE\bfseries}
\titleformat*{\subsection}{\Large\bfseries}
\titleformat*{\subsubsection}{\large\bfseries}
\titleformat*{\paragraph}{\large\bfseries}
\titleformat*{\subparagraph}{\large\bfseries}
\usepackage{bm}
\usepackage{newtxmath,newtxtext}
\usepackage{latexsym}
\usepackage[latin1]{inputenc}
\usepackage{graphicx}
\usepackage{float}
\usepackage{booktabs}
\usepackage{dcolumn}
\usepackage{ragged2e}
\usepackage{hyperref}
\hypersetup{colorlinks,citecolor=blue}
\hypersetup{colorlinks=true,linkcolor=blue,filecolor=magenta,urlcolor=blue}
\usepackage{xcolor}
\usepackage{marvosym}
\usepackage{orcidlink}
\usepackage{epsfig}
\usepackage{cancel, ulem}
\usepackage{subcaption}
\usepackage{caption}
\usepackage{multirow}
\newcommand{\m}{\mathring}
\newcommand{\be}{\begin{equation}}
\newcommand{\ee}{\end{equation}}
\newcommand{\bea}{\begin{eqnarray}}
\newcommand{\eea}{\end{eqnarray}}
\newcommand{\eeas}{\end{eqnarray*}}
\newcommand{\beas}{\begin{eqnarray*}}

\def\jnl@style{\it}
\def\aaref@jnl#1{{\jnl@style#1}}

\def\aaref@jnl#1{{\jnl@style#1}}

\def\aj{\aaref@jnl{AJ}}                   
\def\apj{\aaref@jnl{ApJ}}                 
\def\apjl{\aaref@jnl{ApJ}}                
\def\apjs{\aaref@jnl{ApJS}}               
\def\apss{\aaref@jnl{Ap\&SS}}             
\def\aap{\aaref@jnl{A\&A}}                
\def\aapr{\aaref@jnl{A\&A~Rev.}}          
\def\aaps{\aaref@jnl{A\&AS}}              
\def\mnras{\aaref@jnl{Mon.~Not.~Roy.~Astron.~Soc.}}             
\def\prd{\aaref@jnl{Phys.~Rev.~D}}        
\def\prc{\aaref@jnl{Phys.~Rev.~C}}  
\def\prl{\aaref@jnl{Phys.~Rev.~Lett.}}    
\def\qjras{\aaref@jnl{QJRAS}}             
\def\skytel{\aaref@jnl{S\&T}}             
\def\ssr{\aaref@jnl{Space~Sci.~Rev.}}     
\def\zap{\aaref@jnl{ZAp}}                 
\def\nat{\aaref@jnl{Nature}}              
\def\aplett{\aaref@jnl{Astrophys.~Lett.}} 
\def\apspr{\aaref@jnl{Astrophys.~Space~Phys.~Res.}} 
\def\physrep{\aaref@jnl{Phys.~Rep.}}      
\def\physscr{\aaref@jnl{Phys.~Scr}}       
\def\commat{\aaref@jnl{Comm.~Math.~Phys.}}              
\def\science{\aaref@jnl{Science}}               
\def\cqg{\aaref@jnl{Classical Quant.~Grav.}}            
\def\jpcs{\aaref@jnl{JPCS}}                                     
\def\ijmpd{\aaref@jnl{Int.~J.~Mod.~Phys.~D}}                    
\def\grg{\aaref@jnl{Gen.~Relat.~Gravit.}}               
\def\rpp{\aaref@jnl{Rep.~Prog.~Phys.}}          
\def\npa{\aaref@jnl{Nucl.~Phys.~A}}        
\def\lrr{\aaref@jnl{Living Rev.~Rel.}}                   
\def\jcap{\aaref@jnl{J.~Cosmology Astropart.~Phys.}}    
\def\rmp{\aaref@jnl{Rev.~Mod.~Phys.}}   
\def\epjc{\aaref@jnl{Eur.~Phys.~J.~C}} 
\def\plb{\aaref@jnl{~Phy.~Lett.~B}} 
\def\mpla{\aaref@jnl{Mod.~Phy.~Lett.~A}} 
\def\arxiv{\aaref@jnl{arxiv.org}}

\allowdisplaybreaks[1]

\begin{document}

\title{ Observational constraints on a non-minimally coupled scalar field in non-metricity gravity with an extra degree of freedom }
\author{Saddam Hussain\orcidlink{0000-0001-6173-6140}}
\email{saddamh@zjut.edu.cn}
\affiliation{Institute for Theoretical Physics and Cosmology, Zhejiang University of Technology, Hangzhou 310023, China}
\author{Ghulam Murtaza\orcidlink{0009-0002-6086-7346}}
\email{ghulammurtaza@1utar.my}
\affiliation{Department of Mathematical and Actuarial Sciences, Universiti Tunku Abdul Rahman, Jalan Sungai Long,
43000 Cheras, Malaysia}
\author{Avik De\textsuperscript{\Letter}\,\orcidlink{0000-0001-6475-3085}}
\email{avikde@um.edu.my}
\affiliation{Institute of Mathematical Sciences, Faculty of Science, Universiti Malaya, 50603 Kuala Lumpur, Malaysia}



\begin{abstract}
We investigate a non-minimally coupled scalar field in non-metricity gravity, adopting a non-trivial connection class in spatially flat FLRW spacetimes. In contrast to the coincident-gauge choice, this connection class activates an additional dynamical degree of freedom (DoF) that modifies the background evolution. We study two representative models, defined by power-law and exponential coupling functions with the exponential potential for scalar field. Confronting both models with recent combined datasets - DESI BAO, two independent supernova compilations, and cosmic chronometers - we find that each reproduces an extended matter-dominated era
followed by a stable late-time de-Sitter attractor, with inferred parameters consistent with the standard $\Lambda$CDM model. A complementary autonomous dynamical-system analysis identifies a sequence of critical points describing successive stiff-fluid, transient matter-dominated, and stable de-Sitter phases, ensuring a viable thermal history of the universe. The observational and dynamical results are mutually consistent and support the cosmological viability of the proposed models.
\end{abstract}

\maketitle
\noindent\textsuperscript{\Letter} Corresponding author. 

\section{Introduction}\label{sec00}
Recent cosmological observations \cite{Riess, Tegmark, Kowalski, Komatsu, Abdalla, Lynch2024} provide strong evidence that the universe is currently experiencing an accelerated expansion. General Relativity (GR) requires the introduction of an additional component in the cosmic energy budget to elaborate on this phenomenon.
This component, commonly referred to as dark energy, is characterized by a negative pressure that drives the accelerated expansion of the universe and is incorporated into the gravitational field equations \cite{Gorini:2004by,Ratra:1987rm,Brax:2017idh,Benaoum:2019kyj}. However, the limitations of GR in explaining certain aspects of the universe have encouraged cosmologists to explore theories that extend beyond it. In this perspective, the observed cosmic acceleration is interpreted as a consequence of additional geometric degrees of freedom introduced through modifications to the Einstein-Hilbert action. Such modifications naturally lead to a class of extended or alternative theories of gravity \cite{Clifton2012}, where the accelerated expansion arises from the underlying geometry itself rather than from an additional dark energy component. One of such simplest extensions is $f(\mathring{R})$ gravity, in which the Ricci scalar $\mathring{R}$ in the Einstein-Hilbert action is replaced by a viable function of $\mathring{R}$ \cite{Felice2010}. Another significant category of modified gravity models is scalar-tensor theories \cite{Brans:1961sx,Faraoni:2004pi,Quiros:2019ktw}. In these frameworks, scalar fields interact with gravity and play a crucial role in shaping the cosmic dynamics. Such theories are widely considered promising alternatives for explaining the late-time accelerated expansion of the universe without the need to introduce exotic dark energy components \cite{Perrotta1999, Boisseau2000}. Scalar fields, initially introduced as inflaton fields responsible for driving the early inflationary phase of the universe, have become key elements in contemporary cosmological models. Beyond their role in the early universe, these fields can also be utilized to establish a possible link between the early time inflationary expansion and the late-time accelerated expansion of the universe \cite{Poulin:2018cxd,Poulin:2023lkg,Odintsov:2016plw,Jaime:2022cho,Elizalde:2010ts,Chaussidon:2025npr,Carloni:2025jlk,DAgostino:2021vvv,Dunsby:2023qpb}.

Brans-Dicke theory \cite{Brans:1961sx} represents one of the earliest and most well-known examples of a scalar-tensor theory of gravity. In this framework, a scalar field is introduced and non-minimally coupled to the gravitational sector, effectively modifying the Lagrangian of GR and allowing the gravitational interaction to vary dynamically. The non-minimal interaction between the scalar curvature and the scalar field naturally arises from quantum corrections to scalar fields propagating in curved spacetime \cite{Birrell1984,Parker2009}, and it is also required within the framework of the renormalization procedure \cite{Parker2009}. Such couplings are further motivated by developments in superstring theory \cite{Maeda1386} and in models of induced gravity \cite{Accetta1985}. Moreover, the simplest inflationary scenario based on a minimally coupled scalar field is no longer strongly supported by current observational data \cite{Collaboration2014,Martin2014}. This therefore highlights the need to explore more general frameworks that extend beyond the minimally coupled scalar field paradigm. The model in which the non-minimal coupling vanishes has been studied in detail by Faraoni \cite{Faraoni201}. 

If we go beyond the traditional Levi-Civita geometry, the restrictions of zero torsion and zero non-metricity can be relaxed. Under this broader perspective, the metric teleparallel formulation adopts a connection that has torsion while remaining free of curvature, giving rise to the theory known as $f(\mathbb{T})$ gravity \cite{bahamondefT}. Another alternative is provided by the symmetric teleparallel framework, where the geometry is characterized by non-metricity while both curvature and torsion vanish. This formulation leads to a different line of research known as $f(Q)$ gravity \cite{heisen2024}. With the growing success of the $f(Q)$ theory in exploring gravitational physics \cite{Jensko:2024bee, Alwan:2024lng, Ayuso:2021vtj, DAmbrosio:2021zpm,De2023viability,Kolhatkar2026, Ferreira:2023awf, Dimakis:2025jrl, Dialektopoulos:2025ihc}, a novel extension was introduced in \cite{intro1} which provides an interesting and promising avenue. In this approach, a scalar field $\phi$ interacts with the non-metricity scalar $Q$ through a non-minimal coupling, opening up new possibilities for modified gravity models, called the scalar non-metricity theories of gravity. In the subsequent paper \cite{laur2024}, the authors demonstrated that the standard cosmological epochs can remain stable under certain conditions, although the possibility of finite-time singularities cannot be excluded. A Brans-Dicke formulation within the context of non-metricity gravity was studied in \cite{BD024}, where exact cosmological solutions were found to retain invariant physical properties under conformal transformations between the Jordan and Einstein frames. This work also provided the first analytic solution for this theory. Building on this, \cite{Giacomini2024} demonstrated that conformal transformations do not alter the equilibrium points or the accelerated cosmological solutions, thereby confirming a direct mapping between these two frames, and in \cite{Valcarcel2022}, exact scalarized spherically symmetric solutions were studied. In \cite{Subramaniam2025contast}, scalar perturbations for this framework were derived, yielding modified growth equations that closely reproduce $\Lambda$CDM behavior while enabling constraints from large-scale structure observations. Furthermore, an extended formulation of scalar-tensor non-metricity gravity has been developed by incorporating the boundary term into the action, leading to modified dynamical equations and richer cosmological behavior, as discussed in \cite{Murtaza2026boundary}.

Within the FLRW spacetime, the symmetric teleparallel formulation allows for four different classes of affine connection that respect the underlying spacetime symmetries. In the case of a spatially flat universe, three such connection classes exist. A fundamental characteristic of symmetric teleparallel gravity is the availability of a special coordinate frame where the affine connection identically disappears. In this frame, the covariant derivative becomes equivalent to the usual partial derivative, enabling a clear distinction between inertial contributions and genuine gravitational effects. The connection associated with this coordinate choice is termed the coincident gauge, while other possible connections are classified as non-coincident gauges. The selection of the affine connection is a key ingredient in constructing the theory. Different choices of connection can result in gravitational models that are dynamically distinct, despite being formulated within the same symmetric teleparallel framework. In particular, when the coincident gauge is adopted, the background equations of symmetric teleparallel gravity coincide with those of the metric teleparallel equivalent gravity. As a result, this choice does not introduce any new cosmological dynamics at the background level. To reveal new phenomenological features inherent to symmetric teleparallel gravity and to fully investigate its dynamical structure, it is therefore necessary to move beyond the coincident gauge \cite{Guzman2024noncoin,De2025bexact,Shabani2023noncoin,Subramaniam2023noncoin,Murtaza:2025gme}.

Within $f(Q)$ gravity, which can be recovered as a limiting case of scalar non-metricity theory, the field equations corresponding to the second connection class can be described in terms of scalar fields, leading to a cosmological framework that exhibits quintom-like behavior \cite{Basilakos:2025olm}. The observational analyses presented in \cite{Paliathanasis:2025hjw} indicate that the cosmological scenario associated with the second connection class can challenge the standard $\Lambda$CDM model, as it provides a better fit to the observational datasets. Furthermore, in \cite{Ren2024}, a data-driven reconstruction of the dynamical connection function was performed, demonstrating that the inclusion of this dynamical component improves the fit to observations and is favored over both $\Lambda$CDM and standard $f(Q)$ models according to the AIC and BIC criteria. In scalar non-metricity theory, the additional degree of freedom has also proven fruitful in describing the entire thermal history of the universe \cite{De2025,Murtaza2025steep,Murtaza2025bcurvature}. Motivated by these results, in the present work we focus on the second (non-coincident) connection class within scalar non-metricity gravity in a flat FLRW spacetime. We perform a comprehensive observational analysis and further validate the results through a detailed dynamical system analysis of two representative models by adopting specific forms for the coupling functions and the dynamical time function $\gamma(t)$.

Phase-space analysis provides a powerful framework for investigating the global dynamics of cosmological models by recasting the field equations into an autonomous system, in which different cosmic epochs, such as inflation, radiation domination, matter domination, and late-time acceleration, emerge as critical points. The stability properties of these points determine whether the corresponding phases are stable, transient, or sensitive to initial conditions, while quantities such as the effective equation of state, density parameters, and deceleration parameter establish a direct connection between the mathematical structure and the physical evolution of the universe. Owing to these advantages, Dynamical System Analysis (DSA) has become a widely used tool in cosmology \cite{colley,sadam1,sadam2}. In $f(Q)$ gravity beyond the coincident gauge, DSA has revealed stable de-Sitter attractors, scaling solutions, and $\Lambda$CDM-like evolution across different connection classes and geometries \cite{andfq,Paliathanasis:2023raj,saiakt}, while related studies have further explored the cosmological dynamics \cite{fQBI,Shabani2023a,shabani2023b}. Similarly, in scalar--tensor non-metricity gravity, dynamical analyses have shown that suitable couplings and potential functions can generate different cosmological epochs, including a matter-dominated era and future stable de-Sitter attractors, while also providing insights into the early evolution of the universe \cite{steep,curvature,cqg}. These studies generally constrain the model parameter space based on the nature of the critical points and their stability properties; hence, they are largely restricted to analytical phase-space investigations without confronting the models with observational data. Therefore, in the present work, we complement the dynamical system analysis by adopting power-law and exponential forms for the coupling and potential functions and subsequently confronting the models with a broad range of observational datasets, namely Cosmic Chronometer (CC) data, Baryon Acoustic Oscillation (BAO) measurements from the Dark Energy Spectroscopic Instrument (DESI) Release II, and different Type Ia Supernova samples, including Pantheon+ and the Dark Energy Survey (DES), thereby providing a more comprehensive and robust assessment of their cosmological viability.

The arrangement of the paper is as follows: in Section \ref{sec1}, we outline the mathematical basis of symmetric teleparallel gravity and derive the corresponding field equations within the scalar-tensor theory based on non-metricity. Section \ref{sec3} is devoted to examining the cosmological implications of the second connection class of the theory.  Sections \ref{sec5} and \ref{result} present a comprehensive observational analysis for two specific models along with their corresponding results. In Section \ref{sec6}, we perform a DSA by expressing the cosmological equations in terms of a closed autonomous system. Finally, in  \ref{sec7}, we present our key findings.

\section{Scalar non-metricity gravity theory} \label{sec1}
In symmetric teleparallel gravity, one works with a flat (curvature-free) and torsion-free affine connection
$\Gamma^\lambda{}_{\mu\nu} = \mathring{\Gamma}^\lambda{}_{\mu\nu}+L^\lambda{}_{\mu\nu}$, where
$\mathring{\Gamma}^\lambda{}_{\mu\nu}$ is the Levi-Civita connection and the disformation
$L^\lambda{}_{\mu\nu}$ is built from the non-metricity tensor
$Q_{\lambda\mu\nu}\equiv\nabla_\lambda g_{\mu\nu}\neq0$.
The non-metricity scalar is $Q = Q_{\alpha\beta\gamma}P^{\alpha\beta\gamma}$, with the superpotential
$P^\lambda{}_{\mu\nu}$ defined in the standard way~\cite{intro1}.
We consider the scalar-tensor extension
\begin{equation}
S = \int d^4x\sqrt{-g}\,
\Bigl[\tfrac{1}{2\kappa}f(\phi)Q - h(\phi)\nabla^\alpha\phi\nabla_\alpha\phi - U(\phi)
+ \mathcal{L}_m\Bigr] ,
\label{eqn:action}
\end{equation}
where $f(\phi)$ is the non-minimal coupling, $h(\phi)$ controls the kinetic term and $U(\phi)$ is the scalar potential. Furthermore, $\kappa = 8\pi G = M_{\rm Pl}^{-2}$ denotes the inverse square of the reduced Planck mass.
In the present analysis we take $h(\phi)=\text{constant}$ (this is always possible through a field redefinition,
see e.g.~\cite{cqg}).

Varying the action with respect to the metric, the scalar field $\phi$, and the affine connection yields the corresponding field
equations~\cite{intro1,laur1}:
\begin{align}
\kappa T_{\mu\nu} = f \m G_{\mu\nu}
+ 2f' P^\lambda{}_{\mu\nu}\nabla_\lambda\phi
- h\Bigl(\nabla_\mu\phi\nabla_\nu\phi - \frac12 g_{\mu\nu}\nabla^\alpha\phi\nabla_\alpha\phi\Bigr)
+ \frac12 U g_{\mu\nu}, \label{eqn:metric}\\
f' Q + h'\nabla^\alpha\phi\nabla_\alpha\phi + 2h\m\nabla^\alpha\m\nabla_\alpha \phi - U' = 0, \label{eqn:scalar}\\
(\nabla_\mu - \tilde L_\mu)(\nabla_\nu - \tilde L_\nu)\bigl(4f P^{\mu\nu}{}_\lambda + \kappa\Delta_\lambda{}^{\mu\nu}\bigr) &= 0,
\label{eqn:connection}
\end{align}

where $\m G_{\mu\nu}$ is the Einstein tensor, primes denote $\phi$ derivatives, and
$\Delta_\lambda{}^{\mu\nu}$ is the hypermomentum.
Equation~\eqref{eqn:metric} can be written as $f \m G_{\mu\nu}=\kappa T^{\mathrm{eff}}_{\mu\nu}$,
with an effective dark-energy component sourced by the scalar field.
Finally, note that these three field equations together produce the usual energy-momentum conservation
$\mathring{\nabla}_\mu T^\mu{}_\nu = 0$, provided the hypermomentum vanishes or is conserved independently~\cite{hyper}. 
Interested readers may refer to Refs.~\cite{intro1,laur1} for the detailed mathematical derivation.

\section{Cosmology with Connection Class~II}\label{sec3}
We adopt the spatially flat FLRW metric
\begin{equation}\label{ds:RW}
ds^2 = -dt^2 + a^2(t)\delta_{ij}dx^i dx^j ,
\end{equation}
with Hubble rate $H=\dot a/a$.
A complete classification of torsion-free and curvature-free affine connections compatible
with the FLRW isometries was obtained in~\cite{FLRW/connection}. For the sake of completeness, we offer the classifications in Appendix \ref{classification}.  
Three inequivalent classes exist; here, we focus on Connection Class~II, which introduces a
non-vanishing function $\gamma(t)$ and keeps the gravitational dynamics distinguishable from the metric-teleparallel counterpart. The corresponding non-metricity scalar is derived as
\begin{equation}
Q = 3\bigl(-2H^2 + 3H\gamma + \dot\gamma\bigr) .
\label{eqn:QclassII}
\end{equation}
The metric field equation of scalar non-metricity gravity \eqref{eqn:metric} produces
the corresponding modified Friedmann equations
\begin{align}
\kappa p_m &= f\bigl(-2\dot H - 3H^2\bigr)
            + \frac12(3\gamma - 4H)\dot f
            - \frac12 h \dot\phi^2 + \frac12 U , \label{peq}\\
\kappa\rho_m &= 3H^2 f + \frac32\gamma\dot f
               - \frac12 h \dot\phi^2 - \frac12 U , \label{rhoeq}
\end{align}
The scalar field equation \eqref{eqn:scalar} yields
\begin{align}
\bigl(-6H^2 + 3\dot\gamma + 9H\gamma\bigr)f'
               - h'\dot\phi^2 - 2h(\ddot\phi + 3H\dot\phi) - U' = 0 , \label{scalareq}
\end{align}
The connection field equation \eqref{eqn:connection} yields
\begin{align}
3\gamma\bigl(3\dot f H + \ddot f\bigr) = 0 . \label{connection}
\end{align}
The continuity equation for the matter density becomes
\begin{align}\label{conti_eq}\kappa\dot{\rho}_m+3\kappa H(\rho_m+p_m)=\frac32\gamma\left[  3\dot f H +\ddot f \right]=0.
\end{align}

		\section{Data Analysis}\label{sec5}
	
	In this section, we confront the model with the latest datasets. To achieve this, we first rewrite the entire set of equations of motion in terms of dimensionless dynamical variables, such that the second-order derivatives appearing in the field equations are converted into first-order differential equations, which can be easily implemented in a python pipeline. The dimensionless dynamical variables are:
	\begin{equation}
		x = \frac{\sqrt{\kappa} \dot{\phi}}{\sqrt{6} H_0}, \ y = \frac{\kappa U}{3 H_0^2},  \ \tilde{\Omega}_{\rm m} = \frac{\kappa \rho_{m}}{3 H_0^2}, \ \hbar = \frac{H}{H_0}, \ \alpha_1 = \frac{\kappa U_0}{H_0^2}, \  \alpha_2 = \alpha f_0 \ ,
	\end{equation}
	where $\alpha_{1,2}$ and $H_0$ are dimensionless constants.
    To solve the field equations, we specify the functional forms of the potential of the field, the non-minimal coupling function \(f(\phi)\), and the {dynamical function} \(\gamma(t)\), and then based on these choices, we divide the current scenario into two submodels: Model I and Model II.
	\subsubsection{Model-I}
	For the first case, we choose a power-law function for the potential of the field, and exponential functions for the coupling and dynamical functions:
	\begin{equation*}
		U(\phi) = U_0 \left(\sqrt{\kappa} \phi\right)^{-n}, \ f(\phi) = f_0 \exp(\alpha \sqrt{\kappa} \phi), \ \gamma(N = \ln a) = \exp(\beta N)\ . 
	\end{equation*}
Here, $\gamma$ is solely a time-dependent function. Therefore, for convenience, we choose an exponential function that depends on the re-parameterized time variable $N = \ln a = -\ln (1+z)$, where $z$ denotes the redshift. The algebraic constraint equation is given by:
	\begin{equation}
		3H^2f = \kappa \rho_m - 
		\frac32\gamma\dot f/\sqrt{\kappa} + 
		\kappa \frac12h\dot \phi^2     +\kappa \frac{U}{2}\,.
	\end{equation}
    Here, we re-express Eq.~\eqref{rhoeq} and introduce $\kappa$ to ensure dimensional consistency. In the previous section, the equations were presented without explicitly displaying $\kappa$. However, in this section, we reintroduce $\kappa$ so that the redefined dynamical variables remain dimensionless and all equations in the autonomous system retain a consistent form.
	In terms of dimensionless dynamical variables, this equation reads 
	\begin{equation}
		\hbar^2 f = \tilde{\Omega}_m  - \frac{3}{2} \frac{\gamma \dot{f}}{\sqrt{\kappa} 3 H_0^2} + h x^2 + y/2\ .
	\end{equation}
	This further simplifies to 
	\begin{equation}
		\hbar^2 f = \tilde{\Omega}_m  - \frac{3}{2} \frac{\gamma \alpha f \sqrt{6}x}{3 H_0 \sqrt{\kappa}} + h x^2 + y/2 \ ,
	\end{equation}
where $f = \frac{\alpha_2}{\alpha} \exp\left( \alpha \left(\frac{3y}{\alpha_1}\right)^{-\frac{1}{n}}\right)$.\footnote{Note that this expression depends on the form of the potential. For a power-law type potential, the parameter \(\sqrt{\kappa} \phi = \left(\frac{3 y H_0^2}{\kappa U_0}\right)^{-1/n}\). Throughout the analysis, we fix $h = 1$ and $\kappa = 1$; nevertheless, we explicitly include them in the equations wherever necessary.}
	 The field equation is,
	\begin{equation}
		f_{, \phi} \left(- 6 H^2/\kappa + 3(\dot{\gamma} + 3 H \gamma)/\kappa^{3/2}\right) - h_{,\phi} \dot{\phi}^2 - 2 h (\ddot{\phi} + 3 H \dot{\phi}) - U_{, \phi} = 0 \ .
	\end{equation}
	Using this, we can determine the $\dot{x}/H$ as:
	\begin{equation}
		\frac{dx}{H dt} = -3x - \frac{1}{2\sqrt{6}} \left(\frac{-n}{\hbar} y \left(\frac{\alpha_1}{3y}\right)^{-1/n} \right) - \frac{6 \hbar \alpha_2}{2 \sqrt{6}} \exp\left( \alpha\left(\frac{3  y}{\alpha_1}\right)^{\frac{-1}{n}}\right) + \frac{3 \alpha_2 \exp\left( \alpha\left(\frac{3  y}{\alpha_1}\right)^{\frac{-1}{n}}\right)e^{\beta N}}{2 \sqrt{6} H_0} (\beta +3) \ .
	\end{equation}
Similarly, \(\dot{y}/H\) can be expanded as: 
\begin{equation}
	\frac{dy}{H dt} = \frac{\sqrt{6}x}{\hbar}(-ny) \left(\dfrac{\alpha_1}{3y}\right)^{\frac{-1}{n}} \ .
\end{equation}
The other relevant equation is the continuity equation for the matter density, which is given by
\begin{equation}
	\kappa\dot{\rho_m}+3\kappa H(\rho_m+p_m)
	=\frac32 \frac{\gamma}{\sqrt{\kappa}}\left(  3\dot f H +\ddot f \right).
\end{equation}
The constraint equation for the non-minimal coupling function is given by: 
\begin{equation}
	3\gamma \left(3\dot f H +\ddot f\right)=0,
\end{equation}
Due to this constraint, the matter density becomes minimally coupled to the rest of the sector. Therefore, the energy density scales as $a^{-3}$, and the corresponding dimensionless fractional energy density becomes
\begin{equation}
	\frac{d \tilde{\Omega}_{m}}{H dt} = -3 \tilde{\Omega}_{m} \ .
\end{equation}
Similarly, the Hubble function can be obtained either from the above algebraic equation or from the corresponding differential equation as 
\begin{equation}
	\frac{d \hbar}{dN} = \hbar \frac{\dot{H}}{H^2}\ .
\end{equation}
Here, $\frac{\dot{H}}{H^2}$ is:
\begin{equation}
	f \dot{H} = \frac{-\kappa \rho_{m}}{2} - H \dot{f} + \frac{3 \gamma \dot{f}}{2 \sqrt{\kappa} } - \frac{\kappa \dot{\phi}^2}{2} \ .
\end{equation}
This can further be re-expressed in terms of a dimensionless variable as: 
\begin{equation}
	\frac{\dot{H}}{H^2} = \frac{3}{f \hbar^2} \left(\frac{-1}{2} \tilde{\Omega}_{m} - \frac{\alpha f \sqrt{6} x \hbar }{3} + \frac{\sqrt{6} \gamma \alpha f x}{2 \sqrt{\kappa } H_0} -x^2\right) \ .
\end{equation}
Here, $\gamma = e^{\beta N}$.  The effective equation of state (EoS) of the system becomes:
\begin{equation}
	w_{\rm eff} = \frac{-2}{3} \frac{\dot{H}}{H^2} -1 \ , 
\end{equation}
and the dark energy EoS becomes:
\begin{equation}
	w_{\rm de} = \frac{-2}{3} \left(\frac{3}{f \hbar^2} \left( - \frac{\alpha f \sqrt{6} x \hbar }{3} + \frac{\sqrt{6} \gamma \alpha f x}{2 \sqrt{\kappa } H_0} -x^2\right)\right) -1 \ .
\end{equation}

\subsubsection{Model-II}

In the second case, we retain the same functional form of the {dynamical function} \(\gamma(t)\) as in the previous case; however, we choose a power-law form for the coupling function.
\begin{equation*}
	U(\phi) = U_0 \left(\sqrt{\kappa} \phi\right)^{-n}, \ f(\phi) =  \alpha (\sqrt{\kappa} \phi)^m, \ \gamma(N = \ln a) = \exp(\beta N)\ . 
\end{equation*}
The choice of these functional forms is phenomenological, as no underlying microphysical model has yet been developed to uniquely determine the form of the coupling. Hence, one typically adopts specific functional forms and tests them against observational datasets. The viability of the model is then assessed based on its performance. 

The power-law form can also be interpreted as a truncated expansion of the exponential function. In this way, we can examine the impact of such a non-linear function on the system and evaluate its performance with current datasets. In terms of dynamical variables, the model reads:
\begin{align*}
	f=\alpha\left(\frac{3y}{\alpha_1}\right)^{\frac{-m}{n}},\quad f_{,\phi}=m\alpha \sqrt{\kappa}\left(\frac{3y}{\alpha_1}\right)^{\frac{1-m}{n}}\ .
\end{align*}
The constraint equation becomes:
\begin{equation}
	\hbar^2 f = \tilde{\Omega}_m  - \frac{\sqrt{3}}{\sqrt{2}} \frac{m\gamma \alpha x}{ H_0 \sqrt{\kappa}}\left(\frac{3y}{\alpha_1}\right)^{\frac{1-m}{n}} + h x^2 + y/2 \ .
\end{equation}
Similarly, the Hubble derivative becomes: 
\begin{equation}
	\frac{\dot{H}}{H^2} = \frac{1}{f \hbar^2} \left(\frac{-3}{2} \tilde{\Omega}_{m} - m\alpha \hbar \sqrt{6}x \left(\dfrac{3y}{\alpha_1}\right)^{\frac{1-m}{n}} + \frac{3\sqrt{6} \gamma \alpha m x}{2 \sqrt{\kappa } H_0}\left(\dfrac{3y}{\alpha_1}\right)^{\frac{1-m}{n}} -3x^2\right) \ .
\end{equation}
Corresponding to this model, the first-order evolution equations for the kinetic variable $(\dot{x})$, the potential variable, the fluid equation, and the Hubble parameter are given by:
\begin{eqnarray}
	\frac{dx}{H dt} &=& -3x + \frac{3ny}{2\sqrt{6}\hbar}  \left(\frac{3y}{\alpha_1}\right)^{1/n} + \frac{3m}{\sqrt{6}}\left(\frac{3y}{\alpha_1}\right)^{\frac{1-m}{n}}  \left(\frac{\alpha e^{\beta N}}{2H_0}(\beta+3)-\alpha\hbar\right)  \ , \\
		\frac{dy}{H dt} &=& \frac{\sqrt{6}x}{\hbar}(-ny) \left(\dfrac{3y}{\alpha_1}\right)^{\frac{1}{n}} \ ,\\
		\frac{d \tilde{\Omega}_{m}}{H dt} &=& -3 \tilde{\Omega}_{m} \ ,\\
			\frac{d \hbar}{Hdt} &=&  \frac{1}{f \hbar} \left(\frac{-3}{2} \tilde{\Omega}_{m} - m\alpha \hbar \sqrt{6}x \left(\dfrac{3y}{\alpha_1}\right)^{\frac{1-m}{n}} + \frac{3\sqrt{6} \gamma \alpha m x}{2 \sqrt{\kappa } H_0}\left(\dfrac{3y}{\alpha_1}\right)^{\frac{1-m}{n}} -3x^2\right)\ .
\end{eqnarray}
Here $\frac{d()}{Hdt} \equiv \frac{d()}{dN}$. The equation of the state for the dark energy becomes:
\begin{align}
	\omega_{\rm de}=\frac{3x^2-2\sqrt{6}m \alpha \hbar x \left(\frac{3y}{\alpha_1}\right)^{\frac{1-m}{n}}-\frac{3y}{2}-\frac{3\sqrt{6}m\alpha \gamma x \left(\frac{3y}{\alpha_1}\right)^{\frac{1-m}{n}}}{2\sqrt{\kappa}H_0}}{3x^2+\frac{3y}{2}-\frac{3\sqrt{6}m\alpha \gamma x \left(\frac{3y}{\alpha_1}\right)^{\frac{1-m}{n}}}{2\sqrt{\kappa }H_0}}\ .
\end{align}

\subsection{Observational samples}

$\bullet$ {\bf CC data:} This dataset consists of 32 model-independent measurements of $H(z)$ \cite{Vagnozzi:2020dfn,Jimenez:2001gg,Moresco:2015cya}, with covariance matrix constructed in \cite{Moresco:2012jh,Moresco:2015cya,Moresco:2016mzx} between 15 highly correlated samples.\footnote{The covarinace matrix can be downloaded here: \url{https://gitlab.com/mmoresco/CCcovariance}}. We denote this dataset by `CC'.

$\bullet$ {\bf PP data:} The dataset is based on the Pantheon+ compilation, consisting of 1701 cosmologically viable light-curve measurements from 1550 distinct Type Ia supernovae spanning the redshift range $z \in [0.001,2.26]$  \cite{Brout:2022vxf}. {We filter out the very low-redshift samples and consider the data with $z > 0.01$. As a result, the total number of samples becomes $1657$. We use $m_B$ apparent magnitude column and compute the likelihood by analytically marginalizing over $M_B$, following the prescription in \cite{Goliath:2001af}, as implemented in the \texttt{Cobaya} repository\footnote{Likelihood estimation code: \url{https://github.com/CobayaSampler/cobaya/blob/master/cobaya/likelihoods/sn/pantheonplus.py}.}.} We represent it as ``PP''.

$\bullet$ {\bf DES data:} This dataset incorporates Type Ia supernovae from the Dark Energy Survey (DES-SN5YR), which comprises 1829 distinct SNe \cite{DES:2024jxu,DES:2024hip}. {However, the initially released dataset showed tension with the Pantheon+ samples. Hence, a recently upgraded version of the samples has been released, now referred to as DES-Dovekie. The updated sample consists of 197 low-redshift SNe Ia and 1623 DES likely SNe Ia \cite{DES:2025sig}. The sample is approximately 20\% more robust than the earlier release and exhibits only slight tension with the Pantheon+ sample. The new sample, along with the full covariance matrix, is available in the GitHub repository.\footnote{The updated \href{https://github.com/des-science/DES-SN5YR/tree/main/4_DISTANCES_COVMAT}{DES-Dovekie sample} and likelihood calculation can be found here -- \url{https://github.com/des-science/DES-SN5YR/blob/main/5_COSMOLOGY/Dovekie_cosmosis_likelihood.py}.}}

To estimate the likelihood corresponding to the supernova sample, we compute the luminosity distance for each model as
\begin{equation}
	D_L({z}) = c(1+{z}) \int_0^{{z}} \frac{dz'}{H(z')} \ ,
\end{equation} 
assuming a spatially flat FLRW metric, where \(c\) denotes the speed of light in km/s. Using this, we compute the distance modulus \(\mu\), or apparent magnitude $(m_B)$, as:
\begin{equation}
	\mu \equiv m_{B}-M_{B} = 5\log(D_L/\text{Mpc}) + 25 \  .
\end{equation}
Thus, the parameters are constrained through the chi-square ($\chi^2$) likelihood given as:
\begin{equation}
	-2 \ln (\mathcal{L}) = \chi^2 = \rm \Delta D_{i} \mathcal{C}^{-1}_{ij} \Delta D_j\ ,
\end{equation}
where $\rm \Delta D = \mu_{\rm Obs} - \mu_{\rm Model}$, and $\mathcal{C}^{-1}$ denotes the inverse of the covariance matrix.

$\bullet$ \textbf{DES BAO:} This dataset incorporates Baryon Acoustic Oscillation (BAO) measurements from the Dark Energy Spectroscopic Instrument (DESI) Release II \cite{DESI:2025zgx}, an improved version of DR1 \cite{DESI:2024mwx,DESI:2019jxc,Moon:2023jgl}. The observables correspond to the ratios \(\{D_M/r_d, D_H/r_d, D_V/r_d\}\), where \(D_M\) is the comoving angular diameter distance, \(D_H\) is the Hubble distance, \(D_V\) is the spherically averaged distance, and \(r_d\) is the sound horizon at the drag epoch \cite{eBOSS:2020yzd,DESI:2024mwx}. In this work, we consider \(r_d\) as a free parameter.

Based on the datasets described above, we determine the joint likelihood for three combinations of datasets:
(i) CC + DESBAO,
(ii) CC + DESBAO + PP, and 
(iii) CC + DESBAO + DES.
Thus, the joint likelihood for each sample can be estimated as:
\begin{equation}
	-2 \ln \mathcal{L}_{\rm tot} = \chi^2_{\rm tot}.
\end{equation}
In the considered datasets, both supernova and BAO samples are insufficient to constrain $H_0$, as the supernova samples are degenerate with the absolute magnitude $M_B$, whereas BAO samples are degenerate with the sound horizon $r_d$. Therefore, we anchor these datasets with an external measurement, such as the Hubble data, to obtain constraints on $H_0$, $r_d$, and other model parameters.

We implement the model in a Python pipeline and use a dynamic nested sampler called {\texttt dynesty}, which is efficient in obtaining parameter constraints as well as the evidence corresponding to the model \cite{Sergey:dynesty,2020MNRAS.493.3132S,nandanroy,2004AIPC..735..395S,Higson:dynamics_nested}. We use \texttt{GetDist} \cite{Lewis:2019xzd} to compute marginalized 1D and 2D posteriors from the samples.

\section{Results and Discussion}\label{result}

\begin{table}[t]
	\centering
	\begin{tabular}{l r | l r}
		\hline
		\multicolumn{2}{c}{\bf Model I} & \multicolumn{2}{c}{\bf Model II}\\
		\hline
		Parameters & Range & Parameters & Range\\
		\hline
		\(H_0\)  & [30,100] & \(H_0\)  & $[30,100]$\\
		\(y_0\)  & [0.7, 1.7] & \(10^{y_0}\)  & $[-5.1, 0]$ \\
		\(10^\alpha\) & $[-8, -4]$ & \(\alpha\) & $0.181$ \\
		$10^{\alpha_1}$ & $[-1, 10]$  & ${\alpha_1}$ & $10^{-0.5}$\\
		$n$ & $[0.1, 3]$ & $n$ & $1$\\
		$r_d$ & $[100, 300]$ & $r_d$ & $[100, 300]$\\
		$x_0$ & $10^{-6}$ & $x_0$ & $10^{-6}$\\
		$\beta$  & 1 & $\beta$  & 1\\
		\(f_0\) & 1 & \(f_0\) & 1\\
		
		&& $m$ & $-1.1$\\

		\hline
		\hline
	\end{tabular}
	
\caption{The uniform prior ranges of the model parameters are given inside the square brackets, while the parameters treated as fixed are listed outside the square brackets.}

	\label{tab:prior_range}
\end{table}

{$\bullet$ \bf Model I:} We confront the model with combinations of datasets where CC+DBAO is treated as the baseline, and the supernova samples are kept distinct. Thus, we consider three combinations of datasets: (i) BASE, (ii) BASE+PP, and (iii) BASE+DES. For Model I, we vary the parameters $\{H_0, y_0, 10^{\alpha}, 10^{\alpha_{1}}, n, r_d\}$, and the ranges of variation are tabulated in Tab.~\ref{tab:prior_range}. These ranges are estimated by numerically varying the parameters such that the model produces viable solutions. Beyond these ranges, the model yields complex solutions or the matter density becomes negative, which is physically not viable. 

We fix the parameter \(x_0 \equiv x(N=0) \le 10^{-6}\). For larger values of $x_0$, the scalar field starts to dominate in the past, which in turn affects the matter density. Therefore, to obtain an extended matter-dominated phase, the kinetic part of the field must remain small at the present epoch. Hence, we fix the initial condition corresponding to the kinetic part of the field, as below a certain range it does not affect the dynamics of the system for the considered datasets. 

\begin{figure}[t]
	\includegraphics[scale=0.6]{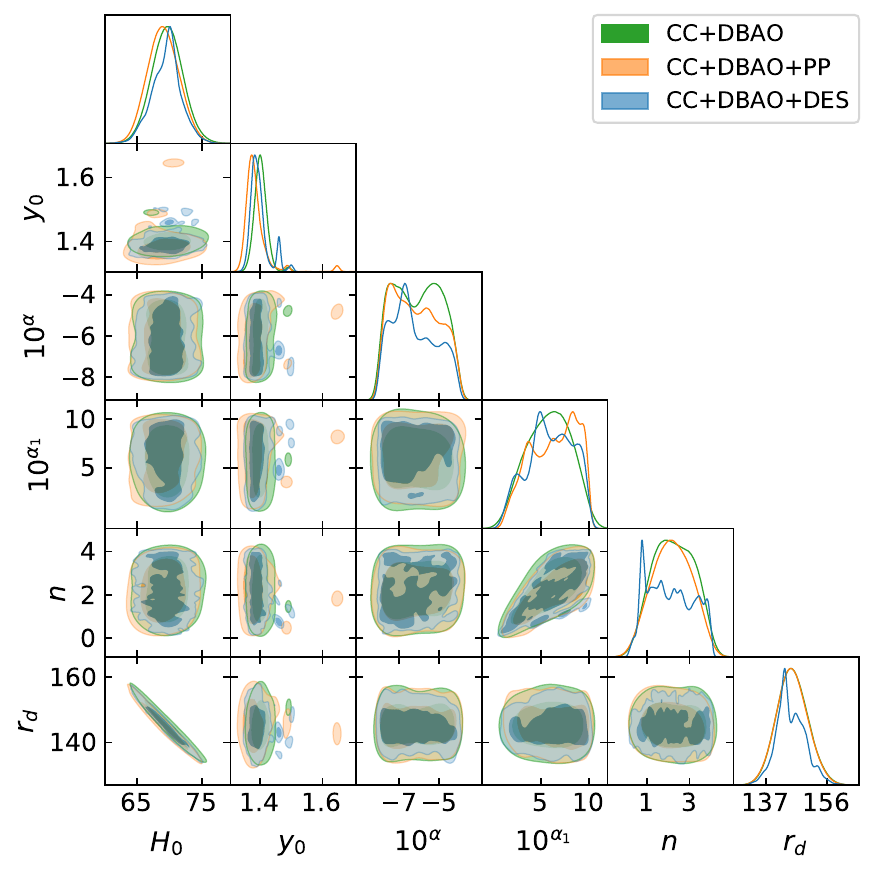}
	\caption{2D Posterior distribution of Model I parameters with various data sets.  }
	\label{fig:corner_model1}
\end{figure}

Similarly, we fix the parameters $\{\beta, f_0\}$, as varying these parameters does not affect the overall evolution of the system, and the current data are not sufficiently sensitive to constrain them. The marginalized posterior distributions of the parameters are shown in Fig.~\ref{fig:corner_model1}, and the corresponding values at $68\%$ confidence level are listed in Tab.~\ref{tab:best_fit_values}. 

The model yields $H_0 \sim 69.2$ km/s/Mpc across all the datasets. Similarly, the sound horizon distance is approximately $145.3$ km/s/Mpc. We have also analyzed the flat $\Lambda$CDM model, which yields very similar values for these parameters.

\begin{table}
	\centering
	\tiny
	\begin{tabular}{l cccccc c l}
		\hline
		\multicolumn{9}{c}{\bf Datasets: CC+DBAO}\\
		\hline
		\hline
		Model & $H_0$ & $\Omega_{\Lambda}$ & $r_d$ & \(y_0\) & $10^{\alpha}$ & $10^{\alpha_1}$ & $n$ & $\log Z$\\
		\hline
		{\bf Model I:} & $69.8 \pm 2.4$ & -- & $145.3\pm 4.9$ &$1.4229^{+0.0035}_{-0.045}$ & $ -6.1 \pm 1.1$ & $5.8 \pm 2.4$ & $2.26\pm 0.96$ & $-24.91$ \\
		{\bf Model II:} & $70.1 \pm 2.4$ & -- & $145.1^{+4.5}_{-5.2}$ & $-4.828 \pm 0.050$ && & & $-25.18$\\
		{\boldmath\bf $\Lambda$CDM:} &$69.8 \pm 2.4$ & $0.6991 \pm 0.0084$& $145.2^{+4.6}_{-5.4}$&&&&& $-23.56$\\
		
		\hline
		\multicolumn{9}{c}{\bf Datasets: CC+DBAO+PP}\\
		\hline
		\hline
		{\bf Model I:} & $69.1 \pm 2.3$ & -- & $145.2\pm 4.9$ & $1.3890^{+0.0077}_{-0.037}$ & $-6.16^{+0.85}_{-1.6}$ & $6.3^{+3.4}_{-3.0}$ & $2.19\pm 0.92$ & $-871.20$\\
		
		{\bf Model II:} & $69.4 \pm 2.5$ & -- & $145.1 \pm 5.1$ & $-4.757 ^{+0.042}_{-0.052}$ && & & $-873.27$\\
		
		{\boldmath\bf $\Lambda$CDM:} &$69.2 \pm 2.4$ & $0.6876 \pm 0.0081$& $145.2\pm 5.0$&&&&& $-869.96$\\
		
		\hline
		\multicolumn{9}{c}{\bf Datasets: CC+DBAO+DES}\\
		\hline
		\hline
		{\bf Model I:} & $69.5 \pm 2.3$ & -- & $144.8 \pm 4.9$ & $1.3995^{+0.0061}_{-0.034}$ & $-6.22^{+0.70}_{-1.7}$ & $6.0^{+3.2}_{-1.8}$ & $2.1\pm 1.1$ & $-846.89$ \\
		
		{\bf Model II:} & $69.6 \pm 2.5$ & -- & $145.2^{+4.8}_{-5.5}$ & $-4.774 \pm 0.044$ && & & $-848.14$\\
		
		{\boldmath\bf $\Lambda$CDM:} &$69.3 \pm 2.4$ & $0.6917 \pm 0.0078$& $145.3\pm 4.9$&&&&& $-845.29$\\
		
		\hline
		\hline
	\end{tabular}
	\caption{The parameter constraint value at 68\% level is obtained by MCMC.}
	\label{tab:best_fit_values}
\end{table}

\begin{figure}[t]
	\includegraphics[scale=0.65]{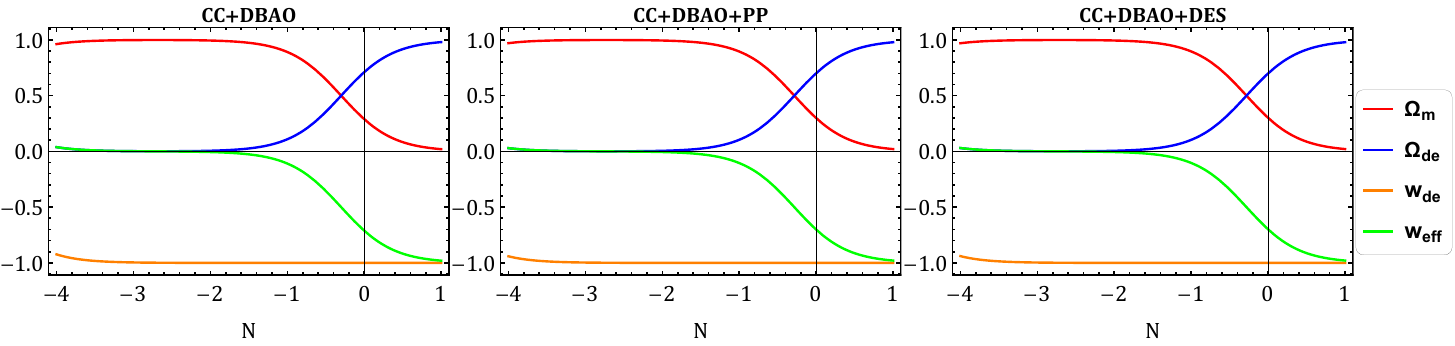}
	\caption{The evolution of fraction energy densities and equation of state against $N= \ln a$ corresponding to Model I for best fit values obtained for distinct data sets.}
	\label{fig:density_evo_model1}
\end{figure}

Corresponding to the parameter values obtained in Tab.~\ref{tab:best_fit_values}, we plot the cosmological quantities such as $\{\Omega_{i} \equiv \frac{\tilde{\Omega}_{i}}{\hbar^2 f}, w_{\rm de}, w_{\rm eff}\}$, 
as functions of the time variable $N$ in Fig.~\ref{fig:density_evo_model1}. 
It can be seen that in the past $(N<0)$, the dark matter density dominates over the effective field density, and the corresponding $ w_{\rm eff}\sim 0$, illustrating a matter-dominated epoch. As evolution proceeds toward the present epoch, the matter density decays while the field density starts to dominate, reaching nearly $70\%$ at the current epoch $N = 0$. The corresponding effective EoS decreases towards negative values and becomes $\sim -0.7$, indicating the accelerated expansion of the universe. 
As the system evolves further into the future $(N>0)$, the field energy density saturates to $1$, and the effective EoS approaches $-1$, indicating a stable de-Sitter phase of the universe. The equation of state of dark energy, $w_{\rm de}$, remains very close to $-1$, mimicking the behavior of a cosmological constant. The evolution of these parameters is very similar across all the datasets.

\begin{figure}[t]
	\includegraphics[scale=0.9]{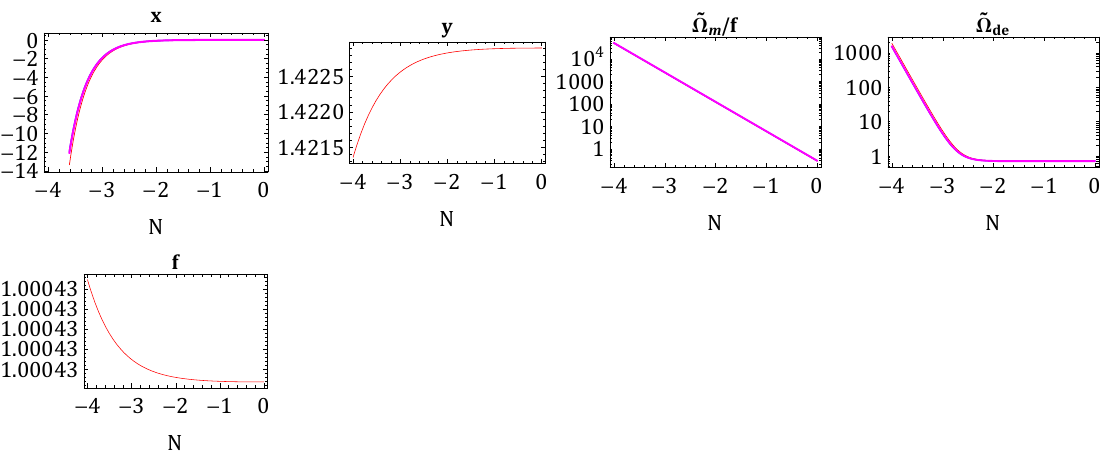}
	\caption{The evolution of dynamical variables and effective densities corresponding to Model I for best fit values obtained for distinct data sets. Here, $\tilde{\Omega}_{\rm de} = 1/f \left( - \frac{3}{2} \frac{\gamma \alpha f \sqrt{6}x}{3 H_0 \sqrt{\kappa}} +  x^2 + y/2\right)$.}
	\label{fig:params_evo_model1}
\end{figure}

Additionally, we plot the dynamical variables and the non-minimal coupling parameter $f$ in Fig.~\ref{fig:params_evo_model1} for the best-fit values obtained for all the datasets. We find that the field parameter $(x)$ increases at higher redshift and becomes extremely small at the current epoch, whereas the parameters $\{y, f\}$ remain nearly constant across redshift. We also plot the density parameters corresponding to the dark matter density $\tilde{\Omega}_{m}$ and the dark energy density $\tilde{\Omega}_{\rm de}$. It is evident that during the matter-dominated epoch, the dark matter density is larger than the dark energy density and remains dominant at higher redshift. However, the dark energy density increases at higher redshift due to the growth of its kinetic function. Nevertheless, the magnitude of the dark energy component remains smaller than that of dark matter, and therefore does not affect the matter-dominated epoch. The non-constant evolution of the dark energy density indicates its dynamical nature, highlighting a significant departure from a cosmological constant.

\begin{figure}[t]
	\includegraphics[scale=0.65]{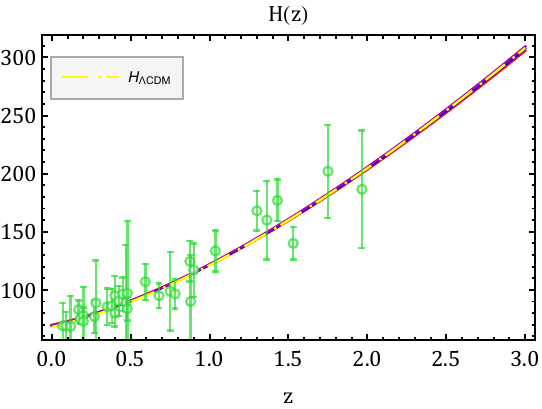}
	\caption{The evolution of Hubble parameter (in solid) corresponding to Model I for best fit values obtained for distinct data sets. The Hubble parameter for $\Lambda$CDM is plotted for $H_0=68.0$, and $\Omega_m= 0.31$.}
	\label{fig:hubble_evo_model1}
\end{figure}

In addition, we plot the Hubble parameter as a function of redshift $z$ and compare it with the CC dataset in Fig.~\ref{fig:hubble_evo_model1}. We also compare the evolution with the $\Lambda$CDM model and find that, at low redshift, both models exhibit a similar behavior.

{$\bullet$ \bf Model II:} The model has a similar number of degrees of freedom as Model I, except for an additional degree of freedom $m$ appearing in the functional form of the non-minimal coupling function $f$. The model consists of several free parameters $\{H_0, y_0, \alpha, \alpha_{1}, n, m, \beta, f_0, r_d\}$, of which we fix the parameters $\{\alpha, \alpha_{1}, n, x_0, \beta, f_0, m\}$. The values of these parameters are tabulated in Tab.~\ref{tab:prior_range}. 
Due to the power-law form of $f$, the model becomes highly non-linear, and often yields either complex solutions or negative dark matter density. For many combinations of parameters, the coupled differential equations become stiff; as a result, varying all parameters simultaneously using the MCMC sampler is not feasible. Therefore, we manually identify and fix suitable values of these parameters. 
We also find alternative parameter values different from those reported in the table; however, these do not lead to qualitatively different behavior. Hence, we choose representative values of the parameters for which the coupled differential equations yield real and physically viable solutions. For these fixed parameters, we vary only $y_0$, which primarily constrains the dark energy density. Although the model suffers from numerical stiffness and several parameters have been fixed, which introduces significant bias into the MCMC analysis, it cannot be regarded as a statistically robust model. Nevertheless, for completeness and to explore its behavior within this reduced parameter space, we present the results obtained from the current model.

\begin{figure}[t]
	\includegraphics[scale=0.6]{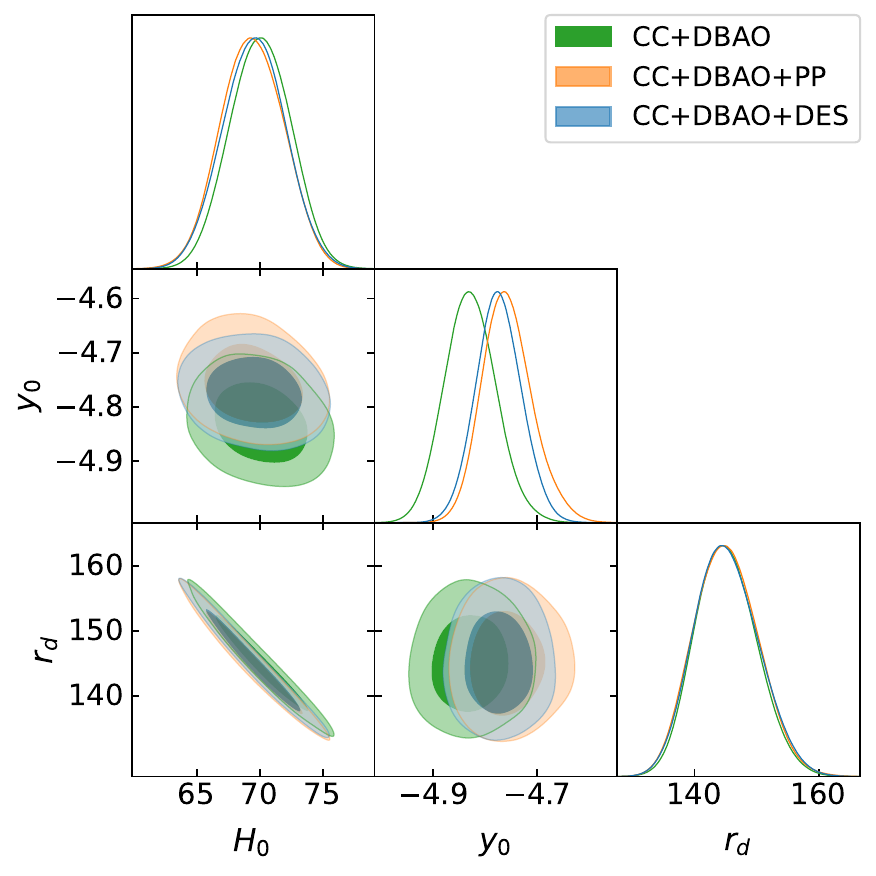}
	\caption{1D-2D Posterior distribution of Model II parameters with various data sets.  }
	\label{fig:corner_model2}
\end{figure}

\begin{figure}[t]
	\includegraphics[scale=0.65]{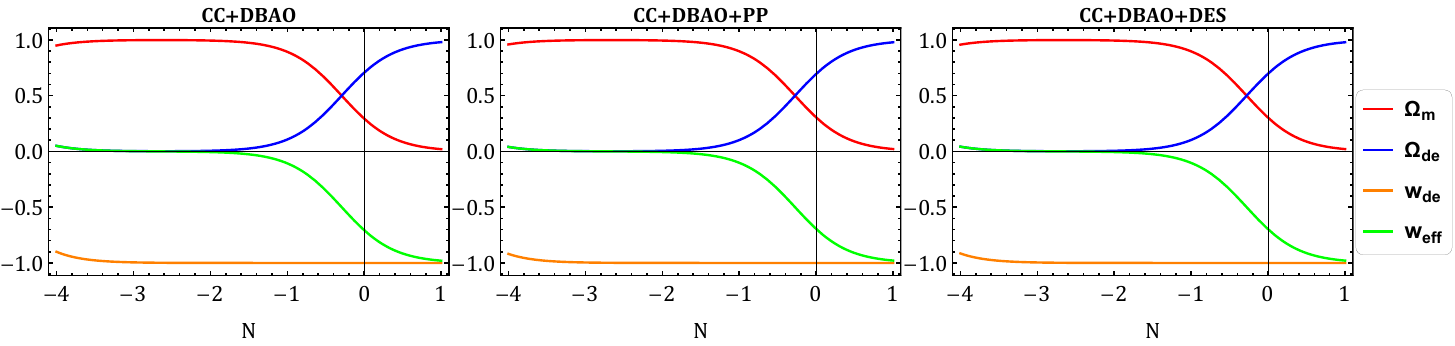}
	\caption{The evolution of fraction energy densities and equation of state against $N= \ln a$ corresponding to Model II for best fit values obtained for distinct data sets.}
	\label{fig:density_evo_model2}
\end{figure}

The posterior distributions of the parameters are shown in Fig.~\ref{fig:corner_model2}, and the corresponding best-fit values at the $68\%$ confidence level are tabulated in Tab.~\ref{tab:best_fit_values}. The model yields $H_0 \sim 70$ km/s/Mpc for BAO observations, while for supernova samples it yields $H_0 \sim 69.5$ km/s/Mpc. The model gives $r_d \sim 145.1$ Mpc. These values are consistent with those of its predecessor model and the $\Lambda$CDM model.

For the best-fit values, we plot the cosmological parameters in Fig.~\ref{fig:density_evo_model2}. We find that the dark matter density for all datasets is nearly $30\%$ at $N=0$, with $w_{\rm eff}\sim -0.65$ while $w_{\rm DE}$ remains close to $-1$, mimicking the cosmological constant. In the past, the model exhibits an extended matter-dominated phase with $w_{\rm eff}\sim 0$. In the future epoch, the dark energy density dominates over the dark matter density; as a result, the effective EoS approaches $-1$, leading to a stable future de-Sitter state.

\begin{figure}[t]
	\includegraphics[scale=0.9]{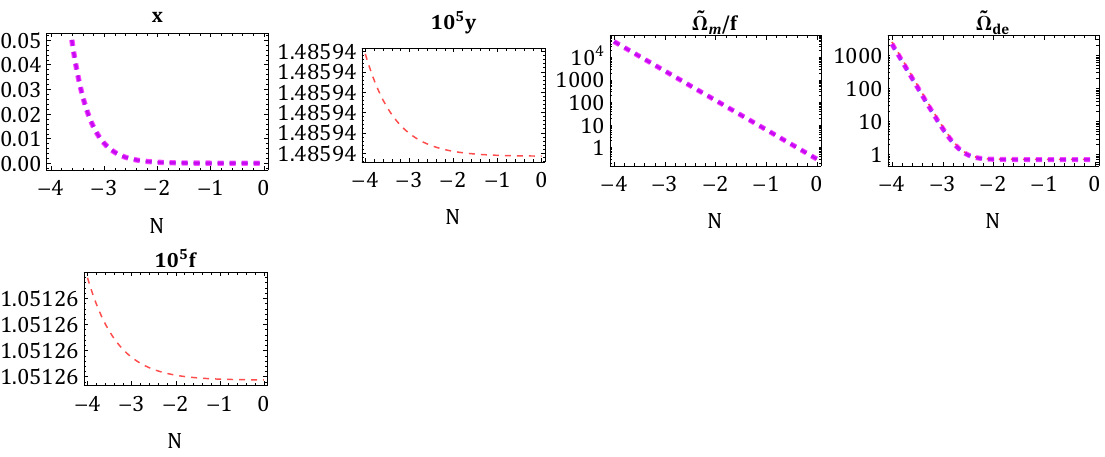}
	\caption{The evolution of dynamical variables and effective densities corresponding to Model II for best fit values obtained for distinct data sets. Here, $\tilde{\Omega}_{\rm de} = 1/f \left( \frac{\sqrt{3}}{\sqrt{2}} \frac{m\gamma \alpha x}{ H_0 \sqrt{\kappa}}\left(\frac{3y}{\alpha_1}\right)^{\frac{1-m}{n}} + x^2 + y/2\right)$.}
	\label{fig:params_evo_model2}
\end{figure}

Additionally, we plot the dynamical variables in Fig.~\ref{fig:params_evo_model2}. Similar to the previous model, $x$ tends to increase at higher redshift, while the other variables $y$ and $f$ remain nearly constant. Here, we plot the rescaled $(y, f)$, as their values are extremely small. A similar behavior of the dark matter and dark energy densities is also observed, consistent with Model I.

\begin{figure}[t]
	\includegraphics[scale=0.65]{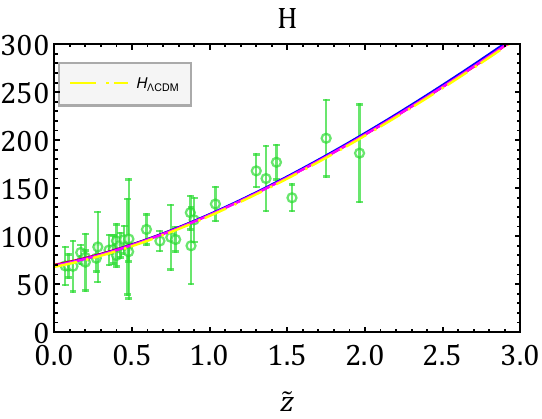}
	\caption{The evolution of Hubble parameter (in solid) corresponding to Model II for best fit values obtained for distinct data sets. The Hubble parameter for $\Lambda$CDM is plotted for $H_0=68.0$, and $\Omega_m= 0.31$.}
	\label{fig:hubble_evo_model2}
\end{figure}

Finally, we plot the Hubble parameter in Fig.~\ref{fig:hubble_evo_model2} for each dataset. The evolution closely matches that of the $\Lambda$CDM model.
 
To compare the model performance with $\Lambda$CDM, we tabulate the $\log Z$ evidence obtained from the nested sampler in Tab.~\ref{tab:best_fit_values}. To perform the comparison, we compute the difference in evidence between a given model and $\Lambda$CDM, defined as $\Delta \log Z_{i\Lambda} \equiv \log Z_{i} - \log Z_{\Lambda}$, where $i$ denotes the model under consideration. 

This difference can be interpreted using the Jeffreys scale: $0 < |\Delta \log Z| < 1$ indicates inconclusive evidence, $1 < |\Delta \log Z| < 2.5$ indicates weak evidence, $2.5 < |\Delta \log Z| < 5$ indicates moderate evidence, and $|\Delta \log Z| > 5$ indicates strong evidence \cite{Escamilla:2024fzq}. For both models, the difference remains negative compared to $\Lambda$CDM, showing a preference for the cosmological constant over the non-minimally coupled models. 
However, for each dataset, Model I outperforms Model II and yields a difference with respect to $\Lambda$CDM of approximately $-1.8$, which suggests that the model considered here cannot be ruled out. Future surveys from the Roman Space Telescope, which will probe supernovae at higher redshift, may play a significant role in distinguishing dynamical dark energy models from the cosmological constant \cite{Hussain:2025vbo,Arora:2025ecj,Hussain:2024yee}.

\section{Cosmological dynamical system Framework}\label{sec6}
In this section, we construct a general autonomous dynamical system framework to study the cosmological dynamics of two particular models. In the present work, we consider a pressureless dust era. In the following, $\overline{( )}$ denotes derivative with respect to $N=\ln a$ or $d/dN$. The phase-space variables are defined from Eq. \eqref{rhoeq} by normalizing with $3fH^2$, leading to the dimensionless variables,
\begin{align}\label{connIIvar}
x=\frac{\dot{\phi}}{\sqrt{6f}H},~~y=\frac{U}{3fH^2},~~z=\frac{\gamma}{H}, (\gamma=e^{\beta N}), ~~\Omega_m=\frac{\kappa \rho_m}{3fH^2}, ~~ \lambda=\frac{U' \sqrt{f}}{U}, ~~ \zeta =\frac{f'}{\sqrt{f}},~~\Gamma=\frac{U'' U}{U'^2},~~\Delta=\frac{f''f}{f'^2},
\end{align}
The constraint equation is written as,
\begin{align} {\label{constr}}
 \Omega_m=1+\sqrt{\frac{3}{2}}xz\zeta-hx^2 -\frac{y}{2}.
\end{align}
Now using the Eq. (\ref{peq}), (\ref{scalareq}), (\ref{connection}), and (\ref{conti_eq}), the autonomous system can be expressed as
\begin{eqnarray}
    \overline{x}&=&-\frac{3x}{2}+\frac{\sqrt{6}x^2 \zeta}{2}+\frac{3hx^3}{2}-\frac{3xy}{4}- \sqrt{6}x^2 \Delta \zeta-\frac{3\sqrt{6}x^2z\zeta}{4},\\
    \overline{y}&=&\sqrt{6}\lambda xy+3y-\frac{3\sqrt{3}yzx \zeta}{\sqrt{2}}+3hx^2 y-\frac{3y^2}{2}+\sqrt{6}xy\zeta,\\
    \overline{z}&=&\frac{3z}{2}+\frac{3hx^2z}{2}-\frac{3yz}{4}+\beta z+\sqrt{6}xz \zeta-\frac{3\sqrt{3}xz^2\zeta}{2\sqrt{2}},\\
\overline{\lambda}&=&\sqrt{6}x\lambda^2(\Gamma-1)+\frac{\sqrt{3}x\lambda \zeta}{\sqrt{2}}, \label{lambda:new}\\
\overline{\zeta}&=&\sqrt{6}x\zeta^2(\Delta -\frac{1}{2}).\label{zeta:eq}
\end{eqnarray}
In this formulation, the autonomous phase space is constructed by taking 
$\{x,y,z,\zeta,\lambda\}$ as the independent dynamical variables, while $\Omega_m$ is not independent but is determined from the Friedmann constraint Eq. (\ref{constr}), which we have used to reduce the dimensionality of the system in our further analysis. Furthermore, the auxiliary quantities $\Gamma$ and $\Delta$ are introduced only to close the dynamical system in the general case, and therefore no additional independent degree of freedom appears once the functional forms are specified in this work. With this structure, the dynamical system is consistently five-dimensional in our analysis, and all critical points are obtained in the reduced autonomous subspace defined by the independent variables. Consequently, the Jacobian matrix used in the stability analysis is constructed only with respect to these independent variables. This ensures that any zero eigenvalues appearing in the spectrum correspond to genuine center directions associated with the intrinsic non-hyperbolic nature of the system, rather than redundancies arising from dependent variables. In terms of these variables, $\frac{\dot{H}}{H^2}$ becomes
\begin{align}
    \frac{\dot{H}}{H^2}=-\frac{3}{2}-\frac{3hx^2}{2}+\frac{3y}{4}-\sqrt{6}x\zeta+\frac{3\sqrt{\frac{3}{2}}xz\zeta}{2}
\end{align}
and the effective Eos parameter is given as,
\begin{align}
    w_{\rm eff}=hx^2-\frac{y}{2}+\frac{x(4-3z)\zeta}{\sqrt{6}}\ .
\end{align}
Also equation of state for the dark energy becomes,
\begin{align}
	\omega_{\rm de}=\frac{\sqrt{6}x \zeta z-\frac{4\sqrt{6}}{3}x\zeta-2hx^2+y}{\sqrt{6}x \zeta z-2hx^2-y}.
\end{align}
In the following subsections, we consider two particular models and perform a thorough stability analysis.

\subsubsection{Model-I}\label{M1}

Hence, for the first case, we choose the following form of these components as: 
\begin{equation*}
	U(\phi) = U_0 \left(\sqrt{\kappa} \phi\right)^{-n}, \ f(\phi) = f_0 \exp(\alpha \sqrt{\kappa} \phi), \ \gamma(N = \ln a) = \exp(\beta N)\ . 
\end{equation*}

For this example we have, $\Delta=1$, and $\Gamma=\frac{1+n}{n}$. So the dynamical system equations can be  expressed as follows,
\begin{eqnarray}
    \overline{x}&=&-\frac{3x}{2}-\frac{\sqrt{6}x^2 \zeta}{2}+\frac{3hx^3}{2}-\frac{3xy}{4}-\frac{3\sqrt{6}x^2z\zeta}{4}, \label{xeq_MI}\\
    \overline{y}&=&\sqrt{6}\lambda xy+3y-\frac{3\sqrt{3}yzx \zeta}{\sqrt{2}}+3hx^2 y-\frac{3y^2}{2}+\sqrt{6}xy\zeta,\\
     \overline{z}&=&\frac{3z}{2}+\frac{3hx^2z}{2}-\frac{3yz}{4}+\beta z+\sqrt{6}xz \zeta-\frac{3\sqrt{3}xz^2\zeta}{2\sqrt{2}},\\
\overline{\lambda}&=&\frac{\sqrt{6}x\lambda^2}{n}+\frac{\sqrt{3}x\lambda \zeta}{\sqrt{2}},\\
\overline{\zeta}&=&\frac{\sqrt{6}x\zeta^2}{2} \label{zetaeq_MI}.
\end{eqnarray}
A summary of the critical points (CPs) of the dynamical system (\ref{xeq_MI})--(\ref{zetaeq_MI}) is provided in Tab. \ref{table1_conll}. The table details the existence conditions of each point, along with $w_{\rm eff}$,  $w_{\rm de}$, $\Omega_m$, their associated eigenvalues, and the corresponding stability characteristics.

The critical points $A_{\pm1}$ correspond to scalar-field dominated solutions in which the kinetic energy of the field completely dominates over both the potential energy and the matter sector. This is evident from $\Omega_m=0$ and $y=0$, while the effective and dark energy equations of state are both equal to unity, $(w_{\rm eff}=w_{\rm de}=1)$. Furthermore, since the additional dynamical variable satisfies $z=0$ at these critical points, the stiff-fluid behavior originates entirely from the scalar-field kinetic energy and not from the extra degree of freedom $\gamma(t)$. Such a state describes a stiff-fluid universe where the pressure equals the energy density, leading to a very rapid dilution of the scalar field energy density as the universe expands. Cosmologically, stiff-fluid solutions are generally associated with the very early stages of cosmic evolution rather than the present accelerated epoch. The existence of two branches, $A_{+1}$ and $A_{-1}$, simply reflects the possibility of the scalar field evolving in opposite directions in field space, although both branches possess identical cosmological properties. The corresponding eigenvalues contain positive directions, indicating that these points are unstable. As a consequence, trajectories in the phase space move away from these solutions under small perturbations. This behavior suggests that the kinetic-dominated epoch can only be transient and may represent an early-time phase from which the universe naturally evolves toward subsequent cosmological eras.

The critical point $A_2$ represents a matter-dominated cosmological phase. At this point, the matter density parameter takes the value $\Omega_m=1$, indicating that the total energy budget of the universe is entirely dominated by pressureless matter, while the scalar field contributions vanish dynamically. Consistent with this interpretation, the effective equation of state is $w_{\rm eff}=0$, which corresponds precisely to the standard dust-dominated era of the universe. The dark energy equation of state becomes undefined at this point because the dark energy density vanishes, leading to an indeterminate expression. This is not a physical inconsistency but simply reflects the absence of a dark energy component. From the dynamical system perspective, $A_2$ is a saddle point since its eigenvalues contain both positive and negative directions. This property is particularly important for cosmological viability, as a realistic matter era should be temporary rather than a final state of evolution. The saddle nature of $A_2$ allows the universe to spend a significant period in a matter-dominated phase before eventually leaving it and evolving toward a late-time accelerated attractor. Therefore, this point provides the necessary intermediate epoch connecting the early universe to the present accelerated expansion.

\begin{figure}[htbp]
    \centering
    
    \begin{subfigure}[t]{0.32\linewidth}
        \centering
        \includegraphics[width=\linewidth]{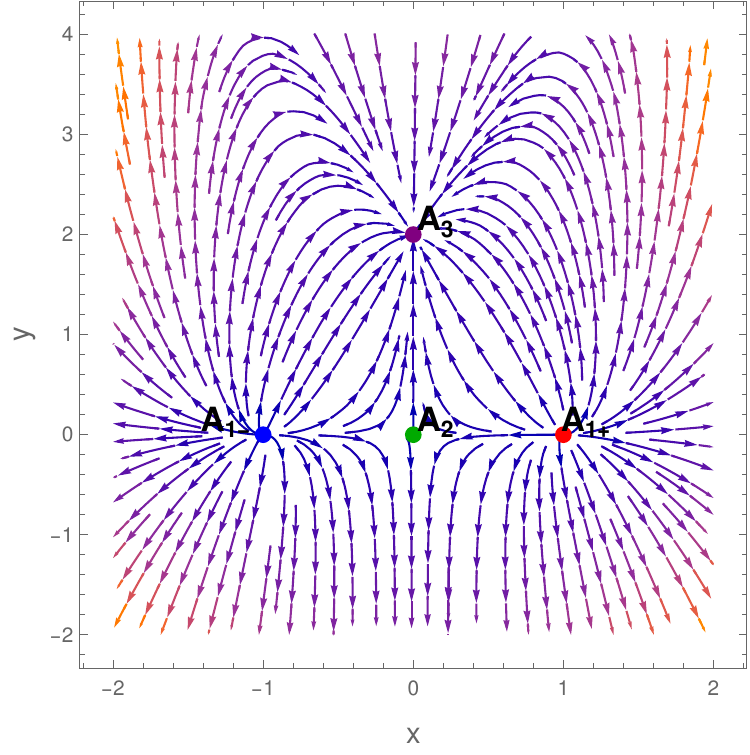}
        \caption{Phase-space portrait for $h=1$.}
        \label{fig1:M1}
    \end{subfigure}
    \hfill
    \begin{subfigure}[t]{0.32\linewidth}
        \centering
        \includegraphics[width=\linewidth]{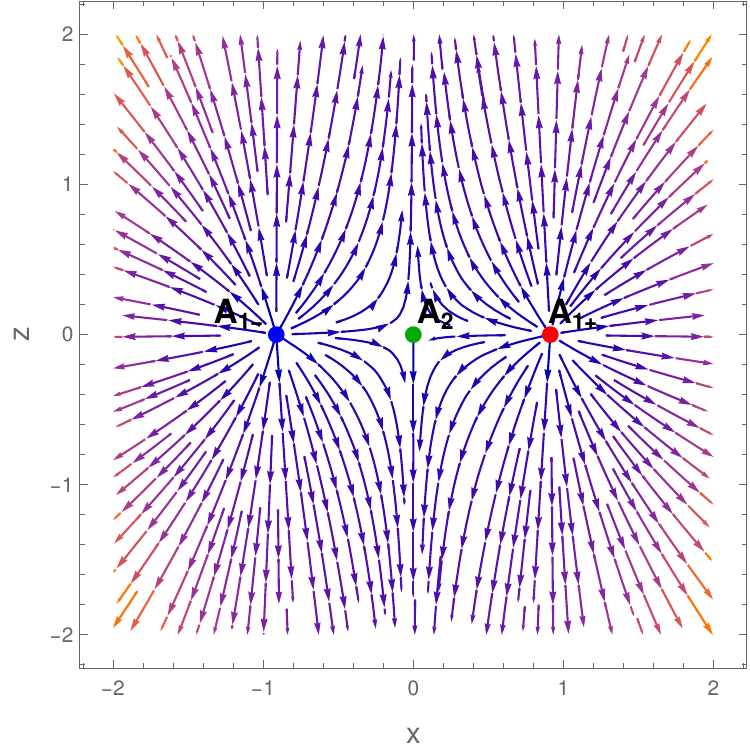}
        \caption{Phase-space portrait  for $h=1.2$ and $\beta=1$.}
        \label{fig2:M1}
    \end{subfigure}
    \hfill
    \begin{subfigure}[t]{0.32\linewidth}
        \centering
        \includegraphics[width=\linewidth]{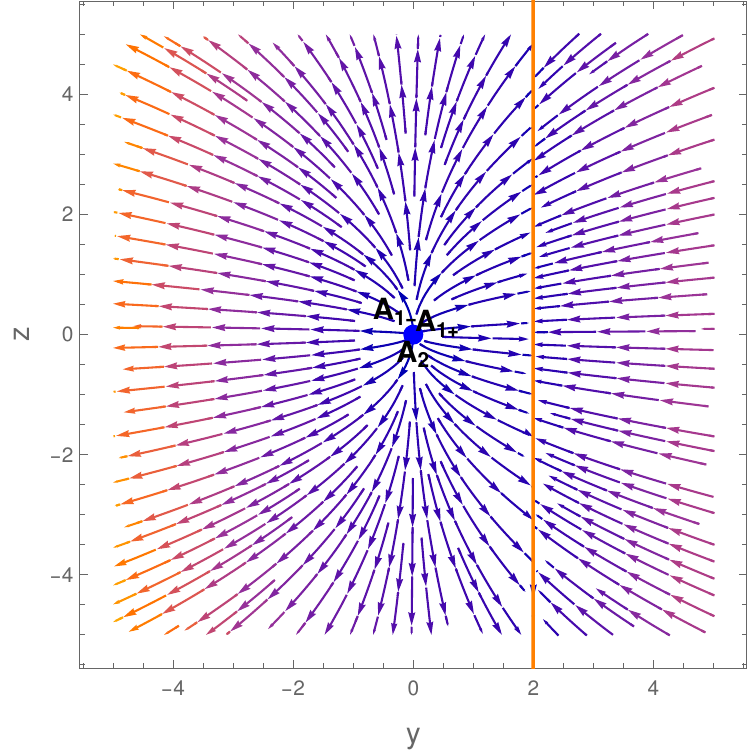}
        \caption{Phase-space portrait  for $\beta=1$.}
        \label{fig3:M1}
    \end{subfigure}

    \caption{The phase-space portraits corresponding to Model-I assuming different values of model parameters.}
    \label{case1_fig1}
\end{figure}

The critical point $A_3$ corresponds to a family of solutions and represents a dark-energy dominated de-Sitter universe, and provides the most cosmologically relevant phase of the model. At this point, the matter density parameter vanishes, $\Omega_m=0$, while both the effective and dark energy equations of state take the value $(w_{\rm eff}=w_{\rm de}=-1)$. These characteristics identify $A_3$ as a de-Sitter universe undergoing exponential expansion driven entirely by the dark energy sector. Although the linear stability analysis yields three zero eigenvalues and therefore classifies the point as non-hyperbolic. However, the center manifold analysis (Appendix \ref{A3_stability}) confirms the stable behavior of this point. Consequently, $A_3$ behaves as a stable late-time attractor of the cosmological dynamics. As $A_3$ is not an isolated critical point but a family of critical points parameterized by the arbitrary constants $z_c$, $\lambda_c$, and $\zeta_c$. Consequently, the late-time attractor of the system is not a single point in phase space. Because of the stable nature, trajectories are attracted toward this manifold, while the particular critical point approached depends on the initial conditions. Therefore, trajectories do not converge to a unique point but to a member of the $A_3$ family with the same asymptotic cosmological properties. In the $(y,z)$ phase-space projection, this family appears as a continuous line at $y=2$, reflecting the arbitrariness of the z-coordinate, as illustrated in Fig.~\ref{fig3:M1}. Taken together, the phase-space analysis suggests a cosmological evolution in which the universe can emerge from an early kinetic-dominated phase, pass through a matter-dominated era, and finally settle into the stable de-Sitter state represented by $A_3$.

\begin{table}[tbp]
\centering
\resizebox{\textwidth}{!}{%
\begin{tabular}{cccccccc}
\hline
\textbf{Critical point}
& $(x,y,z,\lambda,\zeta)$
& \textbf{Existence}
& $\mathbf{\Omega_{\rm m}}$
& $\mathbf{w_{\rm eff}}$
& $\mathbf{w_{\rm de}}$
& \textbf{Eigenvalues}
& \textbf{Stability}
\\ \hline

$A_{\pm1}$
& $\left(\pm\frac{1}{\sqrt{h}},0,0,0,0\right)$
& $h>0$
& $0$
& $1$
& $1$
& $\left(0,0,3,6,3+\beta\right)$
& unstable
\\ \hline

$A_2$
& $\left(0,0,0,\lambda,\zeta\right)$
& $\lambda,\zeta$ arbitrary
& $1$
& $0$
& undefined
& $\left(0,0,-\frac32,3,\frac12(3+2\beta)\right)$
& saddle
\\ \hline

$A_3$
& $\left(0,2,z_c,\lambda_c,\zeta_c\right)$
& $z_c,\lambda_c,\zeta_c$ arbitrary
& $0$
& $-1$
& $-1$
& $\left(0,0,0,-3,-3\right)$
& non-hyperbolic
\\ \hline
\hline

\end{tabular}%
}
\caption{Critical points and their physical properties.}
\label{table1_conll}
\end{table}

\subsubsection{Model-II}

For the second case, we choose the following form of these components as: 
\begin{equation*}
	U(\phi) = U_0 \left(\sqrt{\kappa} \phi\right)^{-n}, \ f(\phi) =  \alpha (\sqrt{\kappa} \phi)^m, \ \gamma(N = \ln a) = \exp(\beta N)\ . 
\end{equation*}
For this model we have, $\Delta=\frac{m-1}{m}$, and $\Gamma=\frac{1+n}{n}$. The dynamical system can be expressed as
\begin{eqnarray}
    \overline{x}&=&-\frac{3x}{2}-\frac{\sqrt{6}x^2 \zeta}{2}+\frac{3hx^3}{2}-\frac{3xy}{4}-\frac{\sqrt{6}x^2\zeta(m-1)}{m}-\frac{3\sqrt{6}x^2z\zeta}{4},\label{xeq_MII}\\
    \overline{y}&=&\sqrt{6}\lambda xy+3y-\frac{3\sqrt{3}yzx \zeta}{\sqrt{2}}+3hx^2 y-\frac{3y^2}{2}+\sqrt{6}xy\zeta,\\
    \overline{z}&=&\frac{3z}{2}+\frac{3hx^2z}{2}-\frac{3yz}{4}+\beta z+\sqrt{6}xz \zeta-\frac{3\sqrt{3}xz^2\zeta}{2\sqrt{2}},\\
\overline{\lambda}&=&\frac{\sqrt{6}x\lambda^2}{n}+\frac{\sqrt{3}x\lambda \zeta}{\sqrt{2}},\\
\overline{\zeta}&=&\frac{\sqrt{6}x\zeta^2(m-2)}{2m}.\label{zetaeq_MII}
\end{eqnarray}
The properties of the critical points of the dynamical system (\ref{xeq_MII})--(\ref{zetaeq_MII}) are concisely provided in Tab. \ref{table3_conll}, including their existence conditions, stability behavior, and corresponding cosmological parameters $w_{\rm eff}$, $w_{\rm de}$ and $\Omega_m$.

The critical points $B_{\pm1}$ describe solutions in which the dynamics of the universe are entirely governed by the kinetic part of the scalar field.  The corresponding values $(w_{\rm eff}=w_{\rm de}=1)$ indicate that the scalar field behaves as a stiff fluid and is characterized by a rapidly decreasing energy density phase of the universe. From a cosmological perspective, these solutions are more naturally associated with the primordial universe, where kinetic effects may temporarily dominate the evolution. The positive eigenvalues appearing in the stability spectrum reveal that these points are unstable. Consequently, $B_{\pm1}$ are best interpreted as early-time transient states that can initiate the cosmological evolution before the universe proceeds toward more physically relevant epochs.

The critical point $B_2$ corresponds to the standard matter-dominated stage of cosmic evolution. At this equilibrium state, the matter density parameter reaches unity, showing that pressureless matter completely controls the expansion dynamics. The value of the effective equation of state is $(w_{\rm eff}=0)$, which is the characteristic signature of a dust-filled universe. Dynamically, $B_2$ exhibits saddle behavior, indicating that it is neither a stable endpoint nor a purely unstable state. This property is essential because it allows the universe to remain in the matter-dominated regime for a substantial period before naturally evolving toward a later accelerated epoch. Therefore, $B_2$ serves as the intermediate bridge connecting the early universe to the dark-energy dominated future.

\begin{figure}[htbp]
    \centering
    
    \begin{subfigure}[t]{0.32\linewidth}
        \centering
        \includegraphics[width=\linewidth]{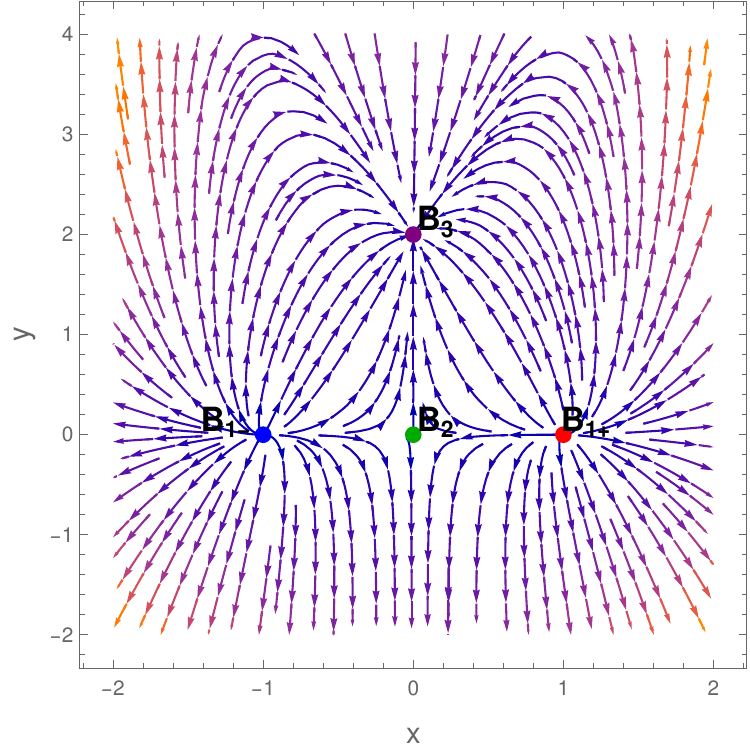}
        \caption{Phase-space portrait for $h=1$.}
        \label{fig1:M2}
    \end{subfigure}
    \hfill
    \begin{subfigure}[t]{0.32\linewidth}
        \centering
        \includegraphics[width=\linewidth]{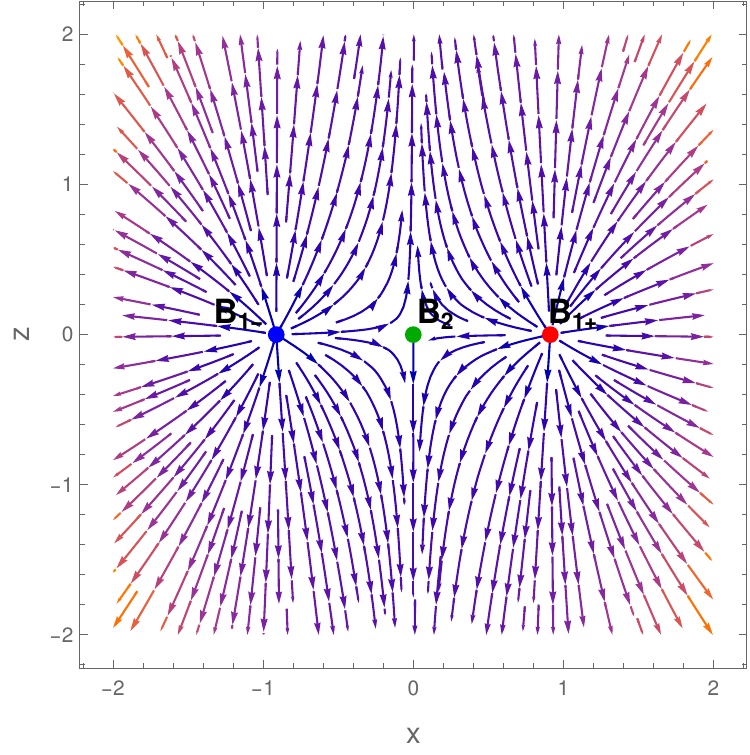}
        \caption{Phase-space portrait for $h=1.2$, $\beta=1$.}
        \label{fig2:M2}
    \end{subfigure}
    \hfill
    \begin{subfigure}[t]{0.32\linewidth}
        \centering
        \includegraphics[width=\linewidth]{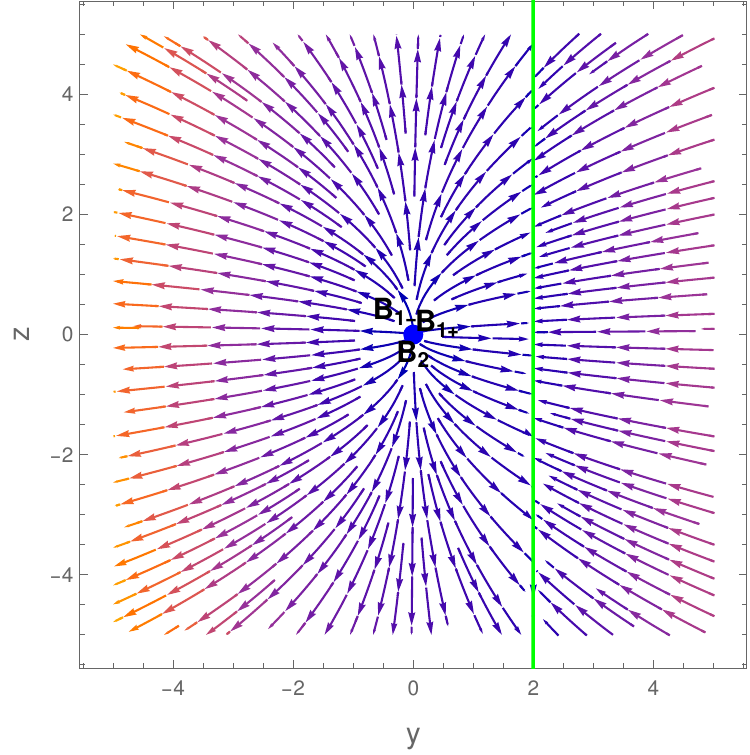}
        \caption{Phase-space portrait for $\beta=1$.}
        \label{fig3:M2}
    \end{subfigure}

    \caption{Phase-space portraits of attractor solution for (\textbf{Model-II}).}
    \label{case2_fig1}
\end{figure}

The critical point $B_3$ represents a family of solutions and corresponds to dark-energy dominated de-Sitter universe. The vanishing matter density parameter together with $ (w_{\rm eff}=w_{\rm de}=-1)$ indicates that the dynamics are entirely driven by an effective cosmological constant like component. In this regime, the expansion rate approaches a constant value, leading to sustained exponential growth of the scale factor. Linear stability analysis fails to identify the stability property of this point because of some zero eigenvalues, but the center manifold analysis in (Appendix \ref{B3_stability}) demonstrates that it is a stable point. This confirms that $B_3$ is a stable attractor of the system and therefore determines the asymptotic fate of the universe. Since modern cosmological observations strongly support a dark-energy dominated epoch with an equation of state close to $-1$, the stability of $B_3$ provides compelling evidence that the model successfully reproduces the observed late-time acceleration and offers a viable description of the future evolution of the universe. Similarly to $A_3$, $B_3$ is also not an isolated critical point but a family of critical points parameterized by the arbitrary constants $z_c$, $\lambda_c$, and $\zeta_c$. Therefore, in the $(y,z)$ phase-space projection, this family appears as a continuous line at $y=2$, reflecting the arbitrariness of the z-coordinate, as illustrated in Fig.~\ref{fig3:M2}.

\begin{table}[tbp]
\centering
\resizebox{\textwidth}{!}{%
\begin{tabular}{cccccccc}
\hline
\textbf{Critical point}
& $(x,y,z,\lambda,\zeta)$
& \textbf{Existence}
& $\mathbf{\Omega_{\rm m}}$
& $\mathbf{w_{\rm eff}}$
& $\mathbf{w_{\rm de}}$
& \textbf{Eigenvalues}
& \textbf{Stability}
\\ \hline
       $B_{\pm1}$ & $\left( \pm\frac{1}{\sqrt{h}},0,0,0,0 \right)$& $h > 0$ &$0$& $1$& $1$& $\left( 0,0,3,6, 3+\beta \right)$ & unstable\\ \hline
         $B_2$ & $\left( 0,0,0,\lambda,\zeta \right)$ &  $\zeta,\lambda=$arbitrary&$1$&$0$& undefined& $\left( 0,0,-\frac{3}{2},3,\frac{1}{2}(3+2 \beta) \right)$ & saddle\\ \hline
          $B_3$ & $\left(0,2,z_c,\lambda_c,\zeta_c \right)$ &$z_c,\zeta_c,\lambda_c=$arbitrary&$0$&$-1$&$-1$& $\left( 0,0,0,-3,-3 \right)$ &non-hyperbolic\\ \hline\hline
          \end{tabular}%
}
\caption{Critical points and their physical properties.}
\label{table3_conll}
\end{table}

\section{Conclusion}\label{sec7}

In this work, we have studied a non-minimally coupled scalar-tensor theory in the framework of non-metricity gravity beyond the coincident gauge. Imposing homogeneity and isotropy on a flat and torsion-free affine connection leads to three classes of cosmological connections characterized by a free time-dependent function $\gamma(t)$. The present analysis has been restricted to the non-trivial connection class II and to an exponential ansatz for $\gamma(t)$, together with two representative choices of the non-minimal coupling function and the power-law potential of the scalar field.

For Model I, corresponding to an exponential coupling function, a full observational analysis using (i) CC+DBAO, (ii) CC+DBAO+PP, and (iii) CC+DBAO+DES datasets shows that the model admits a prolonged matter-dominated epoch followed by a late-time accelerated phase. The dark energy equation of state remains very close to $w_{\rm DE}=-1$, while the inferred values of $H_0$ and $r_d$ are consistent with those obtained in the $\Lambda$ CDM scenario. Although the cosmological evolution closely resembles that of $\Lambda$CDM at the background level, the dark energy sector remains dynamical through the evolution of the scalar field and the non-minimal coupling.

For Model II, which is based on a power-law coupling function, the strong non-linearity of the system and the appearance of numerical stiffness prevent a statistically complete exploration of full parameter space. Consequently, several parameters were fixed to representative values that yield physically viable solutions. The process of such restriction naturally introduces significant bias into the MCMC analysis, and therefore it cannot be regarded as a statistically robust model. Nevertheless, within this restricted parameter space, the model displays a cosmological evolution qualitatively similar to that of Model I, including a matter-dominated era followed by a stable dark-energy dominated de-Sitter phase. The Bayesian evidence slightly favors $\Lambda$CDM over both models, although Model I statistically outperforms Model II and cannot be decisively excluded by the current datasets considered here.

Furthermore, we employ dynamical system analysis to investigate the stability of the cosmological evolution corresponding to the observationally constrained parameter space. Both models admit critical points whose stability is determined using the center manifold theorem for non-hyperbolic points and linear stability analysis for hyperbolic points. For the observationally preferred parameter ranges, both models evolve toward asymptotically stable critical points corresponding to late-time accelerated expansion. These results demonstrate that the models are dynamically stable at the background level and remain observationally viable.

In addition, a full perturbation analysis is left for future work, which may reveal interesting signatures of the time-dependent function $\gamma(t)$ when studying the growth of matter density perturbations using $f\sigma_8$ observations. Although the exponential form of $\gamma(t)$ suppresses its effects at the background level, its role may become more significant in the perturbation sector and could lead to observable deviations from the standard cosmological model.

\begin{acknowledgements}
    S.H. acknowledges the support of National Natural Science Foundation of China under Grants No. W2433018 and No.
		11675143, and the National Key Research and development Program of China under Grant No. 2020YFC2201503. GM acknowledges Universiti Tunku Abdul Rahman Research Fund project IPSR/RMC/UTARRF/2023-C1/A09 provided by Universiti Tunku Abdul Rahman. AD acknowledges the Universiti Malaya BKP-ECRG Grant (Project No. BKP119-2025-ECRG).
\end{acknowledgements}

\appendix

\section{Classification of Connection Classes}\label{classification}
To determine affine connections that are all consistent with the metric in Eq. (\ref{ds:RW}) that provide zero curvature tensor, one imposes two requirements: (i) invariance under six Killing vector fields associated with the metric (\ref{ds:RW}), (ii) $R^{\rho}_{\;\alpha\beta\gamma}=0$. By imposing these along with $\mathcal{L}_X \Gamma^{\beta}_{\;\rho \alpha} = 0$ ($\mathcal{L}$ denoting the Lie derivative) results in four inequivalent FLRW compatible affine connections in symmetric teleparallel gravity: three for flat and one for non-flat cases. These can be written as follows \cite{FLRW/connection}:
\begin{align} \label{eqn:conn}
\Gamma^t{}_{tt}=&C_1, 
	\quad 					\Gamma^t{}_{rr}=\frac{C_2}{1-kr^2}, 
	\quad 					\Gamma^t{}_{\theta\theta}=C_2r^2, 
	\quad						\Gamma^t{}_{\phi\phi}=C_2r^2\sin^2\theta,								\notag\\
\Gamma^r{}_{tr}=&C_3, 
	\quad  	\Gamma^r{}_{rr}=\frac{kr}{1-kr^2}, 
	\quad		\Gamma^r{}_{\theta\theta}=-(1-kr^2)r, 
	\quad		\Gamma^r{}_{\phi\phi}=-(1-kr^2)r\sin^2\theta,												\notag\\
\Gamma^\theta{}_{t\theta}=&C_3, 
	\quad		\Gamma^\theta{}_{r\theta}=\frac1r,
	\quad		\Gamma^\theta{}_{\phi\phi}=-\cos\theta\sin\theta,										\notag\\
\Gamma^\phi{}_{t\phi}=&C_3, 
	\quad 	\Gamma^\phi{}_{r\phi}=\frac1r, 
	\quad 	\Gamma^\phi{}_{\theta\phi}=\cot\theta,
\end{align}
where $C_1$, $C_2$, and $C_3$ are functions of time that are required to meet one of the following conditions: 

\begin{enumerate}\label{casesConn}
\item[(I)] $C_1=\gamma$, $C_2=C_3=0$ and $k=0$, where $\gamma$ is a function of time; or
\item[(II)] $C_1=\gamma+\dfrac{\dot\gamma}\gamma$, $C_2=0$, $C_3=\gamma$ and $k=0$,
             where $\gamma$ is a nonzero function of time; or 
\item[(III)] $C_1=-\dfrac k\gamma-\dfrac{\dot\gamma}{\gamma}$, $C_2=\gamma$, $C_3=-\dfrac k\gamma$ and $k=0,\pm1$,
             where $\gamma$  is a nonzero function of time.
\end{enumerate}

Using (\ref{eqn:conn}), one can get
disformation tensor, superpotential tensor, and non-metricity scalar 
as 
\begin{align}
L^\lambda{}_{\mu\nu}
=&C_1u^\lambda u_\mu u_\nu
+\left(\frac{C_2}{a^2}-H\right)u^\lambda h_{\mu\nu}
+(C_3-H)(-h^\lambda{}_\nu u_\mu-h^\lambda{}_\mu u_\nu)  \label{eqn:L2}\\
2P^\lambda{}_{\mu\nu}
=&\frac12\left(3C_3-3\frac{C_2}{a^2}\right)u^\lambda u_\mu u_\nu
    +\frac12\left(3C_3+\frac{C_2}{a^2}-4H\right)u^\lambda h_{\mu\nu}
    +\frac12(C_1+C_3-H)(-h^\lambda{}_\mu u_\nu-h^\lambda{}_\nu u_\mu) 
    \label{eqn:P2}\\
Q
=&3\left(-2H^2+2\frac k{a^2}+3HC_3+\dot C_3+\frac1{a^2}(HC_2+\dot C_2)\right)\,.
\label{eqn:Q2}
\end{align}
Therefore, the Friedmann equations are derived as
\begin{align}
\kappa p
=&f\left(-2\dot H-3H^2-\frac k{a^2}\right)
    +\frac12\dot f\left(3C_3+\frac{C_2}{a^2}-4H\right)
    -\frac12h\dot \phi^2    +\frac12U\,,\\
\kappa\rho
=&f\left(3H^2+3\dfrac k{a^2}\right)
    +\frac12\dot f\left(3C_3-3\frac{C_2}{a^2}\right)
     -\frac12h\dot \phi^2    -\frac12U\,,\\
0=&\partial_j\phi=\partial_j f
    ; \quad j=1,2,3\,.
\end{align}

\begin{align}
p^{\text{DE}}
=&-\frac1{2f}\left[ 
    \dot f\left(3C_3+\frac{C_2}{a^2}-4H\right)
    -h\dot \phi^2    +U  \right]\,,\\
\rho^{\text{DE}}
=&-\frac1{2f}\left[
    \dot f\left(3C_3-3\frac{C_2}{a^2}\right)
     -h\dot \phi^2    -U \right]\,.
\end{align}
Furthermore, the modified continuity equation follows as
\begin{align}
\kappa\dot{\rho}_m+3\kappa H(\rho_m+p_m)
=\frac32\left[
    \dot f\left(3C_3H-\frac{2\dot C_2+C_2H}{a^2}\right)
    +\ddot f\left(C_3-\frac{C_2}{a^2}\right)\right].
\end{align}

When the hypermomentum tensor is absent, (\ref{eqn:connection}) ensures that the right-hand side of the above expression vanishes.

\section{Stability Analysis using Center Manifold Theory}\label{app1}

Following the frameworks developed by Jack \cite{Carr} and Perko \cite{Perko2013}, we summarize the mathematical basis of center manifold theory. When the Jacobian matrix of an autonomous system at a critical point has one or more zero eigenvalues, linear stability analysis alone is insufficient to determine the system's behavior. In such cases, center manifold theory offers a systematic approach by reducing the system's dimensionality near the critical point. A locally invariant center manifold $W^c$ always exists and passes through the critical point, capturing the essential dynamics in its neighborhood. By studying the flow restricted to $W^c$, one can infer the local stability characteristics of the original system.

\textbf{Theorem}: Let us consider a nonlinear autonomous system in the neighborhood of a fixed point
\begin{align}\label{xeq}
\frac{dx}{dN} &= A x + f(x,y), 
\end{align}
\begin{align}\label{yeq}
\frac{dy}{dN} &= B y + g(x,y).
\end{align}
where $(x,y) \in \mathbb{R}^c \times \mathbb{R}^s$, with $c$ the dimension of $\mathbb{E}^c$ and $s$ the dimension of $\mathbb{E}^s$, and the two functions $f$ and $g$ satisfy
\begin{align}
f(0,0) &= 0,~g(0,0) = 0,\\
Df(0,0) &= 0,~Dg(0,0) = 0.
\end{align}

In the system (\ref{xeq}-\ref{yeq}), $A$ is a matrix of $c \times c$ order and represents eigenvalues having zero real parts, while $B$ is a matrix of $s \times s$ order and denotes the eigenvalues having negative real parts.

\medskip

\textbf{Center Manifold} : 
A geometric space can be regarded as a center manifold for (\ref{xeq}-\ref{yeq}), and locally given as
\begin{equation}
W^c(0) = \{ (x,y) \in \mathbb{R}^c \times \mathbb{R}^s : y = h(x), \, |x|<\delta , \, h(0)=0, \, Dh(0)=0 \},
\end{equation}
provided $\delta$ is sufficiently small and $h(x)$ is a regular function on $\mathbb{R}^s$.

\medskip

The requirements $h(0) = 0$ and $Dh(0) = 0$ ensure that $W^c(0)$ is
tangent to $E^c$ at the critical point $(x,y) = (0,0)$.

\textbf{Existence}
There exists a center manifold associated with system (\ref{xeq}-\ref{yeq}). The dynamics of (\ref{xeq}-\ref{yeq}), when restricted to this center manifold, is described by
\begin{equation}\label{ueq}
\frac{du}{dN} = A u + f(u,h(u)), \quad u \in \mathbb{R}^c \ \text{ sufficiently small}.
\end{equation}

\textbf{Theorem (Stability)}:
If the zero solution of (\ref{ueq}) is stable (whether stable, asymptotically stable, or unstable), then the zero solution of (\ref{xeq}-\ref{yeq}) exhibits the same type of stability. In addition, whenever $(x(N),y(N))$ is a solution of (\ref{xeq}-\ref{yeq}) with $(x(0),y(0))$ sufficiently small, there exists a corresponding solution $u(N)$ of (\ref{ueq}) such that
\begin{align}
x(N) &= u(N) + \mathcal{O}(e^{-\eta N}), \\
y(N) &= h(u(N)) + \mathcal{O}(e^{-\eta N}),
\end{align}
as $N \to \infty$, where $\eta > 0$ is a constant.

where we also used the fact that $y = h(x)$,  must satisfy the quasilinear partial differential equation
\begin{equation}\label{quasi}
\mathcal{N}(h(x)) := Dh(x)[A x + f(x,h(x))] - B h(x) - g(x,h(x)) = 0,
\end{equation}
for it to be the center manifold.  

\subsection{Stability analysis of critical point \texorpdfstring{$A_3$}{A3}}\label{A3_stability}
 We apply the CMT to study the dynamics of the system (\ref{xeq_MI})-(\ref{zetaeq_MI}) near a point $A_3( 0,2,z_c,\lambda_c,\zeta_c)$. Shifting the point $A_3$ to origin via a transformation $x \to x,~y \to y-2,~z \to z-z_c,~\lambda \to \lambda-\lambda_c, \zeta \to \zeta-\zeta_c$ under which the system (\ref{xeq_MI})-(\ref{zetaeq_MI})  becomes
\begin{eqnarray}
\overline{x}&=&\frac{1}{4}x \left( -6 + 6 h x^{2} - 3(2+y) 
- 3\sqrt{6}x\left(z+z_{c}\right) \left(\zeta+\zeta_{c}\right)
- 2\sqrt{6}\,x\left(\zeta+\zeta_{c}\right) 
\right)\ ,\\
\overline{y}&=&\frac{1}{2}(2+y)\left(6+6 h x^{2} - 3(2+y)  + 2\sqrt{6}\,x(\lambda + \lambda_c) + 2\sqrt{6}\,x(\zeta + \zeta_c)-3\sqrt{6}\,x(\zeta + \zeta_c)(z+z_c)\right)\ ,\\
\overline{z}&=&\frac{1}{4}(z+z_c)\left(6+6 h x^{2} - 3(2+y-\frac{4}{3}\beta)  + 4\sqrt{6}\,x(\zeta + \zeta_c)-3\sqrt{6}\,x(\zeta + \zeta_c)(z+z_c)\right)\ ,\\
\overline{\lambda}&=&  \frac{\sqrt{\frac{3}{2}}x(\lambda+\lambda_c)(2\lambda+n\zeta+2\lambda_c+n\zeta_c)}{n}\ ,\\
\overline{\zeta}&=&\sqrt{\tfrac{3}{2}}\,x\,(\zeta+\zeta_{c})^{2}.
\end{eqnarray}
The above system can be written as
\begin{align}
\begin{pmatrix}
\dfrac{dx}{dN} \\
\dfrac{dy}{dN} \\
\dfrac{dz}{dN} \\
\dfrac{d\lambda}{dN}\\
\dfrac{d\zeta}{dN}
\end{pmatrix}
=
\begin{pmatrix}
-3 & 0 & 0 & 0& 0\\
0 & -3 & 0&0& 0 \\
0 & 0 & 0&0& 0\\
0 & 0 & 0&0& 0\\
0 & 0 & 0&0& 0
\end{pmatrix}
\begin{pmatrix}
x \\ y \\ z \\ \lambda \\ \zeta
\end{pmatrix}
+
\begin{pmatrix}
g_1 \\ g_2  \\ f_3 \\ f_4 \\ f_5
\end{pmatrix}
\end{align}

We define the center manifold coordinates as
\begin{equation}\label{cmt2}
\begin{aligned}
 h_{1}(z,\lambda,\zeta) &= a_1z^2 + a_2\lambda^2 + a_3\zeta^2+a_4z\lambda+a_5z\zeta +a_6\lambda \zeta, \\
    h_{2}(z,\lambda,\zeta) &= b_1z^2 + b_2\lambda^2 + b_3\zeta^2+b_4z\lambda+b_5z\zeta +b_6\lambda \zeta.
\end{aligned}
\end{equation}

By solving Eq. (\ref{quasi}) and on comparing like powers of $z^2,~\lambda^2$, $~\zeta^2,~z\lambda,~z \zeta,$ and  $~\lambda\zeta$, we obtain the following coefficients values,
$a_1=0,~a_2=0,~a_3=0~,a_4=0,~a_5=0,~a_6=0,~b_1=\frac{2}{3},~b_2=0,~b_3=0,~b_4=0,~b_5=0,~b_6=0$.\\
Under these, we can rewrite the center manifold coordinates in the form
\begin{equation}
\begin{aligned}
    h_{1}(z,\lambda,\zeta) &= 0+\mathcal{O}(z,\lambda,\zeta), \\
    h_{2}(z,\lambda,\zeta) &= \frac{2}{3}z^2 +\mathcal{O}(z,\lambda,\zeta).
    \end{aligned}
\end{equation}
Finally, the dynamics of the local center manifold are determined by the Eq. (\ref{ueq}), which in our case is given as
 \begin{align}
     \frac{dz}{dN}&=-\frac{z^3}{2}+ higher~ order~ terms,\\
     \frac{d \lambda}{dN}&=0,\\
     \frac{d \zeta}{dN}&=0.
 \end{align}
Although the center directions of $\lambda$ and $\zeta$ are neutral, however the center direction of $z$ is an odd parity order term with a negative sign, implying the stable nature of the point $A_3$.

\subsection{Stability analysis of critical point \texorpdfstring{$B_3$}{B3}}\label{B3_stability}
Transform the point $B_3$ to origin through the following transformation $x \to x,~y \to y-2,~z \to z-z_c,~\lambda \to \lambda-\lambda_c, \zeta \to \zeta-\zeta_c$, so the system (\ref{xeq_MII})-(\ref{zetaeq_MII})  becomes

\begin{align}
\overline{x}
&=\frac{1}{4}x \Big(
 -6 + 6 h x^{2} - 3(2+y)
 - 3\sqrt{6}x(z+z_{c})(\zeta+\zeta_{c})
 -\frac{4\sqrt{6}(m-1)x(\zeta+\zeta_c)}{m}\nonumber\\
&\quad-2\sqrt{6}\,x(\zeta+\zeta_{c})
\Big), \\
\overline{y}
&=\frac{1}{2}(2+y)\Big(
6+6hx^{2}-3(2+y)+2\sqrt{6}\,x(\lambda+\lambda_c)+2\sqrt{6}\,x(\zeta+\zeta_c)-3\sqrt{6}\,x(\zeta+\zeta_c)(z+z_c)\Big), \\
\overline{z}
&=\frac{1}{4}(z+z_c)\Big(
6+6 h x^{2}-3\!\left(2+y-\frac{4}{3}\beta\right)+4\sqrt{6}\,x(\zeta+\zeta_c)
-3\sqrt{6}\,x(\zeta+\zeta_c)(z+z_c)
\Big), \\
\overline{\lambda}
&=\frac{\sqrt{\frac32}\,x(\lambda+\lambda_c)
(2\lambda+n\zeta+2\lambda_c+n\zeta_c)}{n}, \\
\overline{\zeta}
&=\frac{\sqrt{\frac32}(m-2)x(\zeta+\zeta_c)^2}{m}.
\end{align}

which can also be expressed as,
\begin{align}
\begin{pmatrix}
\dfrac{dx}{dN} \\
\dfrac{dy}{dN} \\
\dfrac{dz}{dN} \\
\dfrac{d\lambda}{dN}\\
\dfrac{d\zeta}{dN}
\end{pmatrix}
=
\begin{pmatrix}
-3 & 0 & 0 & 0& 0\\
0 & -3 & 0&0& 0 \\
0 & 0 & 0&0& 0\\
0 & 0 & 0&0& 0\\
0 & 0 & 0&0& 0
\end{pmatrix}
\begin{pmatrix}
x \\ y \\ z \\ \lambda \\ \zeta
\end{pmatrix}
+
\begin{pmatrix}
g_1 \\ g_2  \\ f_3 \\ f_4 \\ f_5
\end{pmatrix}
\end{align}

The coordinates of the center manifold are
\begin{equation}
\begin{aligned}
    h_{1}(z,\lambda,\zeta) &= a_1z^2 + a_2\lambda^2 + a_3\zeta^2+a_4z\lambda+a_5z\zeta +a_6\lambda \zeta, \\
    h_{2}(z,\lambda,\zeta) &= b_1z^2 + b_2\lambda^2 + b_3\zeta^2+b_4z\lambda+b_5z\zeta +b_6\lambda \zeta.
\end{aligned}
\end{equation}

On comparing the same  powers of variables, we get
$a_1=0,~a_2=0,~a_3=0~,a_4=0,~a_5=0,~a_6=0,~b_1=\frac{2}{3},~b_2=0,~b_3=0,~b_4=0,~b_5=0,~b_6=0$.\\
Therefore, the coordinates of the center manifold become,
\begin{equation}
\begin{aligned}
    h_{1}(z,\lambda,\zeta) &= 0+\mathcal{O}(z,\lambda,\zeta), \\
    h_{2}(z,\lambda,\zeta) &= \frac{2}{3}z^2 +\mathcal{O}(z,\lambda,\zeta).
    \end{aligned}
\end{equation}
Hence, the dynamics of the local center manifold are given as
 \begin{align}
     \frac{dz}{dN}&=-\frac{z^3}{2}+ higher~ order~ terms,\\
     \frac{d \lambda}{dN}&=0,\\
     \frac{d \zeta}{dN}&=0.
 \end{align}
implying the stable nature of $B_3$.

\end{document}